\documentclass[preprintnumbers,nofootinbib,noshowpacs,twocolumn,prd]{revtex4}

\usepackage{graphicx}
\usepackage{amsmath,amsthm,amssymb}
\usepackage{mathrsfs,bbm}
\usepackage{cancel}

\makeatletter       

\renewcommand{\p@subsection}{}

\renewcommand{\theequation}{\arabic{equation}}
\makeatother

\newtheorem{theorem}{Theorem}

\newcommand*{\eweakgroup}{\ensuremath{SU(2)_L \times U(1)_Y}}
\newcommand*{\emgroup}{\ensuremath{U(1)_{em}}}
\newcommand*{\CP}{\ensuremath{\mathrm{CP}}}
\newcommand*{\CPviol}{\ensuremath{{\cancel{\mathrm{CP}}}}}
\newcommand*{\CPs}{\ensuremath{\mathrm{CP_s}}}
\newcommand*{\CPg}{\ensuremath{\mathrm{CP_g}}}
\newcommand*{\CPm}{\CP}

\newcommand*{\CPgm}{\CPg}
\newcommand*{\CPviolm}{\CPviol}
\newcommand*{\CPii}{\ensuremath{\text{CP}_g^{(ii)}}}
\newcommand*{\CPiio}{\ensuremath{\overline{\text{CP}}_g^{(ii)}}}
\newcommand*{\unitmatrix}{\mathbbm{1}}

\newcommand*{\abs}[1]{\left\lvert {#1} \right\rvert} 
\newcommand*{\twomat}[1]{\underline{#1}}             
\newcommand*{\tvec}[1]{\ensuremath{\boldsymbol{\mathrm{#1}}}}              
\newcommand*{\fvec}[1]{\ensuremath{\boldsymbol{\mathrm{\tilde{#1}}}}  }    
\newcommand*{\fmat}[1]{\tilde{#1}}                   
\newcommand*{\trans}{\mathrm{T}}                     
\newcommand*{\by}{\!\times\!}                        

\DeclareMathOperator{\trace}{tr}
\DeclareMathOperator{\mRe}{Re}
\DeclareMathOperator{\mIm}{Im}
\DeclareMathOperator{\diag}{diag}

\newcommand*{\hidden}[1]{}

\begin{document}

\preprint{HD-THEP-07-16}

\title{\CP\ Violation in the General\\
       Two-Higgs-Doublet Model: a Geometric View}

\author{M. Maniatis}
    \email[E-mail: ]{M.Maniatis@thphys.uni-heidelberg.de}
\author{A. von Manteuffel}
    \email[E-mail: ]{A.v.Manteuffel@thphys.uni-heidelberg.de}
\author{O. Nachtmann}
    \email[E-mail: ]{O.Nachtmann@thphys.uni-heidelberg.de}

\affiliation{
Institut f\"ur Theoretische Physik, Philosophenweg 16, 69120
Heidelberg, Germany
}


\begin{abstract}
We discuss the \CP\ properties of the potential in the general 
Two-Higgs-Doublet Model (THDM).
This is done in a concise way using real gauge invariant functions
built from the scalar products of the doublet fields. 
The space of these invariant functions, parametrising
the gauge orbits of the Higgs fields, is isomorphic to the
forward light cone and its interior.
\CP\ transformations are shown to correspond to reflections
in the space of the gauge invariant functions.
We consider \CP\ transformations where
no mixing of the Higgs doublets
is taken into account as well as the general
case where the Higgs basis is not fixed.
We present basis independent conditions for explicit
\CP\ violation which may be checked easily
for any THDM potential.
Conditions for spontaneous \CP\ violation, that is \CP\ violation
through the vacuum expectation values of the Higgs fields,
are also derived in a basis independent way.
\end{abstract}


\maketitle
\newpage



\section{Introduction}
\label{sec-intr}

In the Standard Model~(SM) and in many extensions of it
like the Minimal 
Supersymmetric Standard Model~(MSSM)~\cite{Nilles:1983ge, Haber:1984rc} the
electroweak symmetry breaking is accomplished via the
Higgs mechanism.
In the SM,  where one Higgs doublet is introduced,
the Higgs potential is automatically invariant under \CP\ transformations.
Thus, \CP\ violation in the SM only arises via
Yukawa interactions of the Higgs field with the fermions,
that is, through the Kobayashi--Maskawa mechanism~\cite{Kobayashi:1973fv}.

Here we investigate models having the standard weak isospin times
hypercharge (\eweakgroup) gauge group as invariance group and a
Higgs sector with two doublets.
That is, we consider the
general Two-Higgs-Doublet Model (THDM). 
In contrast to the SM, in the THDM
the Higgs potential itself is in general not invariant under \CP\ transformations~\cite{Kobayashi:1973fv}.

The \CP\ properties of the Higgs potential are studied
in the framework of gauge invariant functions, built from all possible
\eweakgroup\ invariant scalar products of Higgs doublets~\cite{Maniatis:2006fs}.
In this approach all invariant scalar products
are replaced by real gauge invariant functions which can
be combined to a four-vector.
In terms of these real gauge invariant functions
a mixing of the Higgs doublets 
corresponds to rotations of the {\em space-like} components
of this four-vector and, as we shall show, 
\CP\ transformations correspond to reflections of the {\em space-like} components.
Thus, constraints for \CP\ invariance can be derived concisely
in this geometric picture.
We also give unambiguous criteria for the occurrence of 
{\em spontaneous} \CP\ violation, where \CP\
violation arises from the vacuum expectation values
of the Higgs doublets, although the Higgs potential itself is \CP\ invariant.

There is much interest in the investigation of an extension
of the Higgs sector for several reasons: 
supersymmetric extensions
require one to have at least two Higgs doublets in order to give masses to
up- and down-type fermions and to keep the theory anomaly free.
Generally, the {\em naturalness} problem arising in the SM
is crucially depending on the Higgs sector.
In~\cite{Barbieri:2005kf} this has been used as a 
motivation to focus on the THDM.
For a recent proposal of THDMs having
a {\em custodial symmetry} see~\cite{Gerard:2007kn}.
Another reason originating from cosmology is that
\CP\ violation is one of the three Sakharov criteria
which have to be fulfilled in order to explain the observed baryon--antibaryon
asymmetry in our Universe through the particle dynamics~\cite{Sakharov:1967dj, Fromme:2006cm}.
In the SM, given the strength of the observed \CP\ violation and the experimental
lower bound on the Higgs mass, 
one cannot explain the baryon excess over
anti-baryons observed in our Universe. For a 
review see for instance~\cite{Bernreuther:2002uj}.
A possible way out of this dilemma is to consider models with an
extended Higgs sector.

There exists already an extensive literature on \CP\ violation in multi-Higgs
and, in particular, two-Higgs-doublet models.
A general discussion of \CP\ transformations in gauge theories was given
in \cite{Grimus:1995zi}.
In~\cite{Lavoura:1994fv, Botella:1994cs} basis independent conditions for spontaneous
\CP\ violation are given for the general THDM.
References~\cite{Davidson:2005cw, Gunion:2005ja} provide an extensive analysis
of the general THDM in terms of invariants with respect to $U(2)$ Higgs
basis changes.
In~\cite{Davidson:2005cw} a proof is given that
the conditions of~\cite{Lavoura:1994fv} for spontaneous \CP\ violation
are sufficient and necessary.
Reference~\cite{Gunion:2005ja} determines the necessary and sufficient conditions
for explicit \CP\ violation in a basis independent way
via the systematic check of potentially complex invariants.
A rather detailed account of \CP\ violation in N-Higgs-doublet models in general
and THDMs in particular was given in \cite{Nishi:2006tg} using gauge invariant
functions.
In~\cite{Ginzburg:2004vp} the Higgs mass squared matrix is considered 
and \CP-conservation conditions are
determined from the possible
mixing of \CP-even and \CP-odd entries in this matrix. 
Reference~\cite{Haber:2006ue} is devoted to spontaneous
symmetry breaking in THDMs, focusing 
critically on the issue if and when the
usual parameter $\tan \beta$ can be considered to be
a truly physical parameter.
A measure for \CP\ violating effects is discussed in~\cite{Khater:2003ym}
for a given Higgs basis and vacuum.
Let us also mention the investigation of the minima structure of THDMs 
in context
with \CP\ violation; see~\cite{Barroso:2007rr}
and references therein.
In \cite{Ivanov:2005hg} the THDM was studied from
a group theoretic point of view.
In \cite{Ivanov:2006yq, Ivanov:2007de} the Minkowski space structure of the $\fvec{K}$-space
(in our notation) was emphasised.
Lorentz transformations were used to diagonalise the term of the
potential~$V$~\eqref{eq-Vfour} quadratic in $\fvec{K}$.
In our present paper we have not used Lorentz transformations in
$\fvec{K}$-space for several reasons.
Lorentz transformations do in general not respect the form of the
kinetic term in the Higgs Lagrangian.
In \cite{Maniatis:2007de} we are interested in the complete theory.
Thus we only consider Higgs-basis transformations which keep
the kinetic term invariant.
There are potentials which are stable in the weak sense
(see section~4 of \cite{Maniatis:2006fs}) and thus completely acceptable
from a physical point of view.
We find examples of such potentials where the term quadratic in
$\fvec{K}$ cannot be diagonalised by a Lorentz
transformation.
In our work we do not exclude these cases from the discussion.
Also we find it generally advantageous to give criteria for properties of a THDM in
a way directly applicable for any given model without assuming a particular choice
for the Higgs-flavour basis.

In our present paper we take up again the question of \CP\ violation in THDMs.
We derive some new results and rederive already known results in a way as
we need it for the companion paper \cite{Maniatis:2007de}.
Indeed, the present paper and \cite{Maniatis:2007de} should be considered as belonging
together and forming one unit.
Our present paper is organised as follows.
In section~\ref{sec-thdm} we briefly recall the definitions of the gauge invariant functions
which provide our framework
to investigate \CP\ properties. 
Then, in section~\ref{sec-explCP}, we classify
the possible types of \CP\ transformations and present
constraints for \CP\ invariance of the potential
in this framework. This is followed
in section~\ref{sec-sponCP} by a discussion of spontaneous \CP\ violation.
The general results are illustrated in section~\ref{sec-exam},
where we discuss two specific models
in the more conventional parametrisation of~\cite{Haber:1993an}.
Section~\ref{sec-conclu} contains our conclusions.
In the respective sections we also compare our findings to those in the literature
mentioned above.
The appendices contain the proofs of two theorems and details for general models
with different types of \CP\ symmetries.


\section{Gauge invariant functions in the general~Two-Higgs-Doublet~Model}
\label{sec-thdm}

We shall use the gauge invariant functions as introduced
in~\cite{Maniatis:2006fs}.
Here we recall the formalism briefly in order to make this work
self-contained.

We denote the two complex Higgs-doublet fields by
\begin{equation}
\label{eq-doubldef}
\varphi_i(x) = \begin{pmatrix} \varphi^+_i(x) \\  \varphi^0_i(x) \end{pmatrix}
\end{equation}
with \mbox{$i = 1, 2$}.  Hence we have eight real scalar degrees of freedom.
The most general \eweakgroup\ invariant Lagrangian for
the THDM can be written as
\begin{equation}
\label{eq-lagr}
\mathscr{L}_{\rm THDM} = \mathscr{L}_{\varphi} + \mathscr{L}_{\rm Yuk} +
\mathscr{L}',
\end{equation}
where the Higgs-boson Lagrangian is given by
\begin{equation}
\label{eq-HiggsL}
\mathscr{L}_{\varphi} = \sum_{i = 1,2}
\left(\mathcal{D}_{\mu} \varphi_i \right)^{\dagger} \left(\mathcal{D}^{\mu}
  \varphi_i \right) - V(\varphi_1, \varphi_2).
\end{equation}
This term replaces the kinetic terms of the Higgs boson and the Higgs
potential in the SM~Lagrangian.
The covariant derivative is
\begin{equation}
\label{eq-covderiva}
\mathcal{D}_{\mu} = \partial_{\mu} + i g  W^a_{\mu} \mathbf{T}_a + i g' B_{\mu}
\mathbf{Y},
\end{equation}
where $\mathbf{T}_a$ and $\mathbf{Y}$ are the generating operators of
weak-isospin and weak-hypercharge transformations.
For the Higgs doublets we have
 \mbox{$\mathbf{T}_a = \tau_a /
2$}, where $\tau_a$ (\mbox{$a = 1,2,3$})
are the Pauli matrices.
We assume both doublets to have weak hypercharge~\mbox{$y = +1/2$}.
By $\mathscr{L}_{\rm Yuk}$ we denote the Yukawa-interaction terms of the Higgs fields
with the fermions.  Finally, $\mathscr{L}'$ contains the terms of the
Lagrangian without Higgs fields.  We do not specify $\mathscr{L}_{\rm Yuk}$
and~$\mathscr{L}'$ here since they are not relevant for our analysis.

We remark that in the~MSSM the two Higgs doublets $H_1$ and $H_2$ carry
hypercharges \mbox{$y = -1/2$} and \mbox{$y = +1/2$}, respectively, whereas
here we use the conventional definition of the~THDM with both doublets
carrying \mbox{$y = +1/2$}.  However, our analysis can be translated to the
other case, see for example (3.1) in~\cite{Gunion:1984yn}, by setting
\begin{equation}
\label{eq-thdmsusytrafo}
\begin{split}
\varphi^\alpha_{1} & = - \epsilon_{\alpha \beta} ( H_1^{\beta} )^{*},\\
\varphi^\alpha_{2} & = H_2^\alpha,
\end{split}
\end{equation}
where $\epsilon$ is given by
\begin{equation}
\label{eq-defeps}
\epsilon = \begin{pmatrix} \phantom{-}0 & 1\phantom{-} \\ -1 & 0\phantom{-} \end
{pmatrix}.
\end{equation}

The most general gauge invariant and renormalisable potential
  \mbox{$V(\varphi_1,\varphi_2)$} for the two Higgs doublets~$\varphi_1$
  and~$\varphi_2$ is a hermitian linear combination of the following terms:
\begin{equation}
\label{eq-potterms}
\varphi_i^{\dagger}\varphi_j,\quad
 \big( \varphi_i^{\dagger}\varphi_j \big)
 \big(\varphi_k^{\dagger}\varphi_l \big),
\end{equation}
where \mbox{$i,j,k,l \in \{ 1, 2\}$}.
It is convenient to discuss the properties of the potential in terms of
gauge invariant expressions.
For this purpose we arrange 
the fields $\varphi_i$~(\ref{eq-doubldef}) in
a $2 \times 2$~matrix~(see (A.2) of~\cite{Maniatis:2006fs})
\begin{equation}
\label{eq-genCPphi}
\phi(x)=
\begin{pmatrix}
\varphi_1^+(x) & \varphi_1^0(x)\\
\varphi_2^+(x) & \varphi_2^0(x)
\end{pmatrix}.
\end{equation}
Similarly, we arrange
the \eweakgroup\ invariant
scalar products
into the hermitian \mbox{$2 \by 2$}~matrix
\begin{equation}
\label{eq-kmat}
\twomat{K}(x) :=
\begin{pmatrix}
  \varphi_1^{\dagger}\varphi_1 & \varphi_2^{\dagger}\varphi_1 \\
  \varphi_1^{\dagger}\varphi_2 & \varphi_2^{\dagger}\varphi_2
\end{pmatrix} = \phi(x) \phi^\dagger(x)
\end{equation}
and consider its decomposition
\begin{equation}
\label{eq-kmatdecomp}
\twomat{K}_{i j}(x) =
 \frac{1}{2}\,\left( K_0(x)\,\delta_{i j} + K_a(x)\,\sigma^a_{i j}\right),
\end{equation}
using the completeness of the Pauli matrices $\sigma^a$ ($a=1,2,3$)
together with the unit matrix.
Here and in the following summation over repeated indices is understood.
Explicitly, (\ref{eq-kmat}) and (\ref{eq-kmatdecomp}) yield
\begin{equation}
\label{eq-phik}
\begin{alignedat}{2}
\varphi_1^{\dagger}\varphi_1 &= (K_0 + K_3)/2, &\quad
\varphi_1^{\dagger}\varphi_2 &= (K_1 + i K_2)/2, \\
\varphi_2^{\dagger}\varphi_2 &= (K_0 - K_3)/2, &
\varphi_2^{\dagger}\varphi_1 &= (K_1 - i K_2)/2\,.
\end{alignedat}
\end{equation}
Thus the four real coefficients defined by the decomposition~(\ref{eq-kmatdecomp})
are given by
\begin{equation}
\label{eq-kdef}
\begin{aligned}
K_0 &= \varphi_1^{\dagger} \varphi_1 + \varphi_2^{\dagger} \varphi_2, &
K_1 &= 2 \mRe \varphi_1^\dagger \varphi_2,\\
K_3 &= \varphi_1^{\dagger} \varphi_1 - \varphi_2^{\dagger} \varphi_2, &
K_2 &= 2 \mIm \varphi_1^\dagger \varphi_2\,.\\
\end{aligned}
\end{equation}
Using the three-vector notation
\begin{equation}
\tvec{K}(x):=
\begin{pmatrix}
K_1(x)\\
K_2(x)\\
K_3(x)
\end{pmatrix},
\end{equation}
the most general potential can be written as follows:
\begin{equation}
\label{eq-vdef}
V = \xi_0 K_0 + \tvec{\xi}^\trans \tvec{K} + \eta_{00} K_0^2 + 2 K_0\, \tvec{\eta}^\trans \tvec{K} 
+ \tvec{K}^\trans E \tvec{K},
\end{equation}
with
\begin{equation}
\tvec{\xi}:=
\begin{pmatrix}
\xi_1\\
\xi_2\\
\xi_3
\end{pmatrix},\quad
\tvec{\eta}:=
\begin{pmatrix}
\eta_1\\
\eta_2\\
\eta_3
\end{pmatrix},\quad
E:=
\begin{pmatrix}
\eta_{11} & \eta_{12} & \eta_{13}\\
\eta_{21} & \eta_{22} & \eta_{23}\\
\eta_{31} & \eta_{32} & \eta_{33}
\end{pmatrix}.
\end{equation}
Here the 14 independent potential parameters $\xi_0$, $\xi_a$, $\eta_{00}$, $\eta_a$
and \mbox{$\eta_{ab}=\eta_{ba}$} are real.

Now we consider a change of basis of the Higgs fields,
$\varphi_i \rightarrow \varphi'_i$, where
\begin{equation}
\label{eq-udef}
\begin{pmatrix} \varphi'_1 \\
                \varphi'_2 \end{pmatrix}
= \begin{pmatrix} U_{11} & U_{12} \\
                  U_{21} & U_{22} \end{pmatrix}
  \begin{pmatrix} \varphi_1 \\
                  \varphi_2 \end{pmatrix} .
\end{equation}
Here
\begin{equation}
U = \begin{pmatrix} U_{11} & U_{12} \\
                    U_{21} & U_{22} \end{pmatrix},
\qquad
U^\dagger U = \unitmatrix_2,
\end{equation}
is a \mbox{$2 \by 2$}~unitary matrix.
With~\eqref{eq-udef} the gauge invariant functions~(\ref{eq-kdef}) transform as
\begin{equation}
\label{eq-biltrafo}
K'_0 = K_0,
\qquad
K'_a = R_{ab}(U) K_b,
\end{equation}
where \mbox{$R_{ab}(U)$} is defined by
\begin{equation}
\label{eq-defrot}
U^\dagger \sigma^a U = R_{ab}(U)\,\sigma^b.
\end{equation}
The matrix~\mbox{$R(U)$} has the properties
\begin{equation}
\label{eq-rprodet}
R^\ast(U)=R(U),
\quad
R^\trans(U)\, R(U) = \unitmatrix_3,
\quad
\det R(U) = 1,
\end{equation}
where $\unitmatrix_3$ denotes the \mbox{$3 \by 3$}~unit matrix.
The transformations fulfill $R(U) \in SO(3)$, that is, they are
proper rotations in $\tvec{K}$-space. 

The Higgs potential~(\ref{eq-vdef}) remains unchanged under
the replacements~(\ref{eq-biltrafo}) if we perform an appropriate transformation
of the parameters of $V$:
\begin{equation}
\label{eq-partrafo}
\begin{alignedat}{2}
\xi'_0 &= \xi_0,  & \tvec{\xi}' &= R(U)\,\tvec{\xi}, \\
\eta'_{00} &= \eta_{00}, &  \tvec{\eta}' &= R(U)\,\tvec{\eta}, \\
 E' &= R(U)\,E\,R^\trans(U).
\end{alignedat}
\end{equation}
Moreover, for every matrix~$R$ with the properties~(\ref{eq-rprodet}),
there is a unitary transformation~(\ref{eq-udef}).
We can therefore diagonalise~$E$, thereby reducing the number of parameters
of~$V$ by three.  The Higgs potential is then determined by only 11 real
parameters.

The matrix~$\twomat{K}(x)$ is positive semi-definite, which
follows immediately from its definition~(\ref{eq-kmat}).
With \mbox{$K_0 = \trace{\twomat{K}}$} and
\mbox{$K_0^2 - \tvec{K}^2 = 4 \det{\twomat{K}}$} this implies
\begin{equation}
\label{eq-kinq}
K_0(x) \geq 0,\qquad K_0(x)^2 - \tvec{K}(x)^2 \geq 0.
\end{equation}
On the other hand, for any given $K_0(x), \tvec{K}(x)$
fulfilling~(\ref{eq-kinq}),
it is possible to find fields $\varphi_i$ obeying~(\ref{eq-kdef}).
Furthermore, all fields obeying~\eqref{eq-kdef} for a given
$K_0(x), \tvec{K}(x)$ form one gauge orbit;
see appendix A of~\cite{Maniatis:2006fs}.

Thus, the functions~$K_0(x), K_a(x)$ parametrise the gauge orbits
and not a unique Higgs-field configuration.
Specifying the domain of the functions~$K_0(x), K_a(x)$ corresponding
to the gauge orbits allows to discuss the potential
directly in the form~(\ref{eq-vdef})
with all gauge degrees of freedom eliminated.
We note that
the gauge orbits of the Higgs fields
of the THDM are parametrised by Minkowski type four-vectors
\begin{equation}
\label{eq-Kfour}
\fvec{K}(x)=
\begin{pmatrix}
K_0(x)\\
\tvec{K}(x)
\end{pmatrix}
\end{equation}
which have to lie on or inside the forward light cone.
This allows us to write the most general 
potential~(\ref{eq-vdef}) in the
concise form~(see (87) and (88) of~\cite{Maniatis:2006fs})
\begin{equation}
\label{eq-Vfour}
V= \fvec{K}^\trans \fvec{\xi} + \fvec{K}^\trans \fmat{E} \fvec{K}\,,
\end{equation}
where
\begin{equation}
\label{eq-fourpar}
\fvec{\xi} = 
\begin{pmatrix}
\xi_0\\
\tvec{\xi}
\end{pmatrix},
\qquad
\fmat{E} =
\begin{pmatrix}
\eta_{00} & \tvec{\eta}^\trans\\
\tvec{\eta} & E
\end{pmatrix}.
\end{equation}
%


\section{\CP~transformations and \CP~invariance~of~the~Lagrangian}
\label{sec-explCP}

\subsection{The standard \CP\ transformation}
\label{sec-standCP}
The standard \CP\ transformation of the gauge fields and the Higgs fields reads
(see for instance~\cite{Nachtmann:1990ta})
\begin{equation}
\begin{split}
\label{eq-simCP-Y}
W^\mu(x) &\xrightarrow{\CPs} -W_\mu^\trans(x'), \phantom{\qquad (i=1,2)}\\
B^\mu(x) &\xrightarrow{\CPs} -B_\mu(x'), \phantom{\qquad (i=1,2)}
\end{split}
\end{equation}
\begin{equation}
\label{eq-simCP}
\varphi_i(x) \xrightarrow{\CPs} \phantom{-}\varphi_i^*(x') \qquad (i=1,2).
\end{equation}
Here we have
\begin{equation}
\left(x^\mu\right)=
\begin{pmatrix}
x^0\\ \tvec{x}
\end{pmatrix},
\quad
\left(x'^\mu\right)=
\begin{pmatrix}
\phantom{-}x^0\\ -\tvec{x}
\end{pmatrix}
\quad
\end{equation}
and
\begin{equation}
W^\mu(x)= W^{\mu a}(x) \frac{1}{2} \tau_a
\end{equation}
is the matrix of the $W$-potentials.
Of course, a discussion of this \CP\ transformation
makes only sense once we have already
chosen a particular basis for the two
Higgs doublets since basis transformations~(\ref{eq-udef})
change~(\ref{eq-simCP}). Such a particular choice
of basis is, indeed, in general required
when the Yukawa term~$\mathscr{L}_{\rm Yuk}$ 
is taken into consideration.
In the MSSM, for instance, one Higgs doublet couples
to the up-type fermions, one to the down type
fermions. This clearly singles out a special
basis. 
Therefore, we have denoted the
\CP\ transformations in~(\ref{eq-simCP-Y})
and~(\ref{eq-simCP}) by \CPs\ for {\em standard}
and {\em special}.

From the definition of our matrix
$\twomat{K}$ and of the
four real coefficients
$K_0$ and $K_a$ it is obvious
that the \CPs\ transformations~(\ref{eq-simCP})
correspond to 
\begin{equation}
\begin{split}
\label{eq-simCP-K}
\twomat{K}(x) &\xrightarrow{\CPs} \twomat{K}^*(x')=\twomat{K}^\trans(x')\,,\\
K_0(x)&\xrightarrow{\CPs}
K_0(x')\,,\\
\begin{pmatrix}
K_1(x)\\ K_2(x)\\ K_3(x)
\end{pmatrix}
&\xrightarrow{\CPs}
\begin{pmatrix}
\phantom{-}K_1(x')\\ -K_2(x')\\ \phantom{-}K_3(x')
\end{pmatrix}.
\end{split}
\end{equation}
That is, the vector $\tvec{K}(x)$ is subjected
to a reflection on the $1$--$3$ plane and a change
of argument $x \rightarrow x'$,
\begin{equation}
\label{eq-rotK}
\tvec{K}(x) \xrightarrow{\CPs} R_2 \tvec{K}(x'),\\
\end{equation}
where
\begin{equation}
\label{eq-R2}
R_2 :=
\begin{pmatrix}
1& 0& 0\\
0& -1& 0\\
0& 0& 1
\end{pmatrix}.
\end{equation}
The potential $V$~(\ref{eq-vdef})
allows for \CPs\ as a symmetry if and only if
it contains no terms linear in $K_2$.
The kinetic term in the Higgs-Lagrangian~(\ref{eq-HiggsL})
is invariant under \CPs\ as defined
in~(\ref{eq-simCP-Y}), (\ref{eq-simCP}). Thus, we have
the following theorem.
\begin{theorem}
\label{theorem-CPs}
The Higgs Lagrangian~(\ref{eq-HiggsL})
with the general potential~(\ref{eq-vdef})
is invariant under the \CPs\ transformation~(\ref{eq-simCP-Y}), (\ref{eq-simCP})
if and only if
\begin{equation}
\label{eq-CPs1}
\xi_2=0,
\quad 
\eta_2=0,
\quad 
\eta_{12}=\eta_{23}=0. 
\end{equation}
Equivalently, we can formulate~(\ref{eq-CPs1})
with the help of the reflection matrix~(\ref{eq-R2})
as
\begin{equation}
\label{eq-CPs1R}
R_2 \tvec{\xi}=\tvec{\xi},
\quad 
R_2 \tvec{\eta}=\tvec{\eta},
\quad 
R_2 E R_2^\trans=E. 
\end{equation}
\end{theorem}


\subsection{Generalised \CP\ transformations}
\label{sec-genCP}

We shall in this paper also consider {\em generalised}
\CP\ transformations of the Higgs fields \cite{Ecker:1987qp}
defined by 
\begin{equation}
\label{eq-genCP}
\varphi_i(x) \xrightarrow{\CPgm} U_{\varphi, ij} \; \varphi_j^*(x'),
\end{equation}
with $i=1,2$ and $U_\varphi = \left(U_{\varphi, ij} \right) \in U(2)$.
That is, the complex conjugation of the Higgs fields is supplemented by 
a basis transformation~(\ref{eq-udef}).
The transformation of the gauge potentials 
stays the same as in~(\ref{eq-simCP-Y}),
\begin{equation}
\begin{split}
\label{eq-genCP-Y}
W^\mu(x) &\xrightarrow{\CPgm} -W_\mu^\trans(x'),\\
B^\mu(x) &\xrightarrow{\CPgm} -B_\mu(x').
\end{split}
\end{equation}

The \CPg\ transformation~(\ref{eq-genCP}) implies for the gauge invariant 
functions~(\ref{eq-kmat}) and (\ref{eq-kmatdecomp})
\begin{equation}
\begin{alignedat}{2}
\label{eq-genCP-K}
\twomat{K}(x)  &\xrightarrow{\CPgm} \; U_\varphi \twomat{K}^*(x')
U_\varphi^\dagger,&\\
K_0(x)  &\xrightarrow{\CPgm}\; K_0(x'),&\\
\tvec{K}(x) &\xrightarrow{\CPgm}\; R(U_\varphi) R_2 \tvec{K}(x'),&
\end{alignedat}
\end{equation}
with $R(U_\varphi) \in SO(3)$ obtained from~(\ref{eq-defrot}) with
$U$ replaced by $U_\varphi$.
That is, \CPg\ induces an {\em improper} rotation $\bar{R}_\varphi$ of
the vector $\tvec{K}$ in addition to the change of argument $x \rightarrow x'$:
\begin{equation}
\label{eq-genCPKvec}
\begin{split}
K_0(x)  &\xrightarrow{\CPgm} K_0(x'),\\
\tvec{K}(x) &\xrightarrow{\CPgm} \bar{R}_\varphi \tvec{K}(x'),
\end{split}
\end{equation}
where 
\begin{equation}
\label{eq-genCPRtilde}
\begin{split}
&\bar{R}_\varphi = R(U_\varphi) R_2,\\
&\bar{R}_\varphi \bar{R}_\varphi^\trans = \unitmatrix_3,\\
&\det \bar{R}_\varphi = \det \left(R(U_\varphi) R_2\right) = -1.
\end{split}
\end{equation}
From the results of section~\ref{sec-thdm} it is
clear that to any improper rotation $\bar{R}_\varphi$ there
is a $U_\varphi \in U(2)$ which, inserted in (\ref{eq-genCP}),
gives (\ref{eq-genCPKvec}) and (\ref{eq-genCPRtilde}).

Thus, asking if the potential $V$~(\ref{eq-vdef})
allows for a \CPg\ symmetry is the same as asking
if it is invariant under some improper
rotation (\ref{eq-genCPKvec}) of the $\tvec{K}$-vectors.
That is, we have invariance under a \CPg\ transformation if
the parameters of $V$~(\ref{eq-vdef}) satisfy
\begin{equation}
\label{eq-genCP1}
\bar{R}_\varphi \tvec{\xi}=\tvec{\xi},
\quad 
\bar{R}_\varphi \tvec{\eta}=\tvec{\eta},
\quad 
\bar{R}_\varphi E \bar{R}_\varphi^\trans=E
\end{equation}
for some improper rotation matrix~$\bar{R}_\varphi$.

We shall study now the effect of a basis 
change~(\ref{eq-udef}) on~$\bar{R}_\varphi$.
For this it is convenient to work with the matrix
$\phi(x)$~(\ref{eq-genCPphi}).
Let the new basis fields be $\varphi_1'(x)$, $\varphi_2'(x)$
and the corresponding matrix
\begin{equation}
\label{eq-genCPchi}
\phi'(x)=
\begin{pmatrix}
\varphi'_1{}^{\!+}(x) & \varphi'_1{}^{\!0}(x)\\
\varphi'_2{}^{\!+}(x) & \varphi'_2{}^{\!0}(x)
\end{pmatrix}=
U \phi(x)
\end{equation}
with $U \in U(2)$. The \CPg\ transformation~(\ref{eq-genCP})
reads
\begin{align}
\label{eq-genCPphi2}
\quad\phi(x) &\xrightarrow{\CPgm} U_\varphi \phi^*(x').\\
\intertext{This implies}
\label{eq-genCPchiU}
\phi'(x) &\xrightarrow{\CPgm} \; U U_\varphi \phi^*(x')\nonumber\\
&\phantom{\xrightarrow{\CPgm}}= U U_\varphi U^{*-1} \phi'{}^{\!*}(x')\\
&\phantom{\xrightarrow{\CPgm}}= U'_\varphi \phi'{}^{\!*}(x'),\nonumber
\end{align}
where
\begin{equation}
\label{eq-genCPdefUchi}
U'_\varphi = U U_\varphi U^{*-1}.
\end{equation}
The transformation of $K'_{0}(x)$ and $\tvec{K}'(x)$ in
the new basis is
\begin{equation}
\begin{alignedat}{1}
\label{eq-genCP-Kchi}
&K'_{0}(x) \xrightarrow{\CPgm}\; K'_{0}(x')\,,\\
&\tvec{K}'(x) \xrightarrow{\CPgm}\; \bar{R}'_\varphi
\tvec{K}'(x')\,,
\end{alignedat}
\end{equation}
with 
\begin{equation}
\label{eq-genCPRtildechi}
\bar{R}'_\varphi = R (U'_\varphi) R_2= R(U) \bar{R}_\varphi R^\trans(U).
\end{equation}
Here $R(U)\in SO(3)$ is the rotation matrix obtained from $U$ according
to~(\ref{eq-defrot}).
Thus, a basis change induces an orthogonal transformation of the improper
rotation matrix~$\bar{R}_\varphi$.

Now we shall consider two successive \CPg\ transformations.
For the gauge potentials and for the gauge invariant
functions we find from~(\ref{eq-genCP-Y}) and (\ref{eq-genCPKvec}):
\begin{equation}
\begin{split}
\label{eq-genCPtwo}
& W^\mu(x) \xrightarrow{\CPgm \circ \CPgm} \;W^\mu(x),\\
& B^\mu(x) \xrightarrow{\CPgm \circ \CPgm} \;B^\mu(x),\\
& K_0(x) \xrightarrow{\CPgm \circ \CPgm} \; K_0(x),\\
& \tvec{K}(x) \xrightarrow{\CPgm \circ \CPgm} \; ( \bar{R}_\varphi )^2 \tvec{K}(x).\\
\end{split}
\end{equation}
Requiring that $\CPgm \circ \CPgm$ gives the unit transformation for
the gauge invariant functions leads to the condition
\begin{equation}
\label{eq-genCPcond}
\bar{R}_\varphi \bar{R}_\varphi =\unitmatrix_3.
\end{equation}
But we also have $\bar{R}_\varphi \bar{R}_\varphi^\trans =\unitmatrix_3$;
see~(\ref{eq-genCPRtilde}). The requirement~(\ref{eq-genCPcond}) thus
means that $\bar{R}_\varphi$ is symmetric
\begin{equation}
\label{eq-genCPsymm}
\bar{R}_\varphi^\trans = \bar{R}_\varphi.
\end{equation}
As a real symmetric matrix it can be diagonalised by an orthogonal
matrix~$R(U)$. That is, we can make a basis change of the
Higgs fields as in~(\ref{eq-genCPchi}) and achieve
\begin{equation}
\label{eq-genCPRchidiag}
\bar{R}'_\varphi = R(U) \bar{R}_\varphi R^\trans(U) = \text{diagonal matrix}.
\end{equation}
Since $\bar{R}'_\varphi$ is an improper rotation it satisfies
$\bar{R}'_\varphi \bar{R}'_\varphi{}^{\!\trans}=\unitmatrix_3$ and
$\det \bar{R}'_\varphi=-1$.
Thus, we have only the possibilities
$\bar{R}'_\varphi = R_1$ or $R_2$ or $R_3$ or $-\unitmatrix_3$.
Here 
\begin{equation}
\begin{split}
\label{eq-genCPreflect}
R_1 &:= \diag(-1, \phantom{-}1, \phantom{-}1),\\
R_2 &:= \diag(\phantom{-}1, -1, \phantom{-}1),\\
R_3 &:= \diag(\phantom{-}1, \phantom{-}1, -1).
\end{split}
\end{equation}
The cases $\bar{R}'_\varphi=R_j$, $j=1,2,3$ are
equivalent by a basis change.
Thus we find the following.

An improper rotation $\bar{R}_\varphi$ satisfying
$\bar{R}_\varphi^2=\unitmatrix_3$ is either
\begin{align}
\label{eq-genCPsum1} 
&(i)\phantom{i} \quad \bar{R}_\varphi=-\unitmatrix_3,\\
\intertext{that is, a point reflection, or orthogonally equivalent to the reflection $R_2$}
\label{eq-genCPsum2}
&(ii) \quad \bar{R}_\varphi= R^\trans(U) R_2 R(U),
\end{align}
that is, a reflection on a plane.

\subsubsection*{\CPg\ transformations of type~$(i)$}

For the case $(i)$, $\bar{R}_\varphi$ as
in~(\ref{eq-genCPsum1}), the \CPg\ transformation
for the fields is obtained from~(\ref{eq-genCPphi2})
by setting $U_\varphi=\epsilon$,
\begin{equation}
\label{eq-genCPphieps}
\phi(x) \xrightarrow{\CPgm} \epsilon \; \phi^*(x'),
\end{equation}
where $\epsilon$ is defined in~(\ref{eq-defeps}).
With this we obtain indeed
\begin{equation}
\label{eq-genCP-Kmateps}
\begin{split}
\twomat{K}(x) &\xrightarrow{\CPgm} 
\epsilon \phi^*(x') \phi^{*\dagger}(x') \epsilon^\trans\\
&\phantom{\xrightarrow{\CPgm}}= \epsilon \twomat{K}^\trans(x') \epsilon^\trans\\
&\phantom{\xrightarrow{\CPgm}}= \frac{1}{2} \big( K_0(x') \unitmatrix_2 - \tvec{K}(x') \tvec{\sigma} \big),\\
\tvec{K}(x) &\xrightarrow{\CPgm} - \tvec{K}(x').
\end{split}
\end{equation}
Note that here $\CPgm \circ \CPgm$ gives the unit transformation
for the Higgs fields only after a suitable gauge transformation.
We have
\begin{equation}
\label{eq-genCPphiepstwo}
\phi(x) \xrightarrow{\CPgm \circ \CPgm} 
\epsilon \left( \epsilon \phi^*(x) \right)^* = -\phi(x).
\end{equation}
A hypercharge gauge transformation
\begin{equation}
U_G=\exp \left( 2 \pi i {\bf Y}\right)
\end{equation}
with ${\bf Y}= \frac{1}{2} \unitmatrix_2$ for the Higgs fields gives 
(see (A.7) of \cite{Maniatis:2006fs})
\begin{equation}
\label{eq-genCPphiY}
\phi(x) \xrightarrow{U_G} \phi(x) U_G^\trans
= \phi(x) (-1).
\end{equation}
Thus, for the case $(i)$, (\ref{eq-genCPsum1}), the
transformation
\begin{equation}
\exp \left( 2 \pi i {\bf Y}\right) \circ \CPgm \circ \CPgm
\end{equation}
is the unit transformation for the Higgs fields and, as
we easily check, also for the gauge potentials.
In appendix~\ref{sec-ap-stability} we show that,
up to gauge transformations, the transformation
of the fields given in~(\ref{eq-genCPphieps}) is the
only possible one giving a \CPg\ transformation
of type $(i)$. We also show there that
(\ref{eq-genCPphieps}) holds in {\em any} basis, again
up to gauge transformations. Thus the
\CPg\ transformations of type $(i)$
have the very interesting, one might even say
aesthetic, property of having the same form
in any Higgs basis.

The invariance conditions for the potential parameters,
(\ref{eq-genCP1}), give us here the following theorem.
\begin{theorem}
\label{theorem-CPgi}
The Higgs boson Lagrangian~(\ref{eq-HiggsL}) with the 
potential~(\ref{eq-vdef}) has the \CPg\ symmetry
(\ref{eq-genCP-Y}), (\ref{eq-genCPKvec})
of type $(i)$, where 
$\bar{R}_\varphi=-\unitmatrix_3$ (see (\ref{eq-genCPsum1})),
if and only if
\begin{equation}
\label{eq-genCPtheorem}
\tvec{\xi}=0 \quad \text{and} \quad \tvec{\eta}=0.
\end{equation}
\end{theorem}
We note that the statements of theorem~\ref{theorem-CPgi} are basis
independent, since the conditions
$\tvec{\xi}=0$ and $\tvec{\eta}=0$ are not affected by
a change of basis. This is a direct consequence of the basis
independence of the form of the \CPg\ transformation
of type~$(i)$.

\subsubsection*{\CPg\ transformations of type~$(ii)$}

For the case
$(ii)$, 
$\bar{R}_\varphi$ as in (\ref{eq-genCPsum2}),
we find that the original \CPg\ transformation
(\ref{eq-genCP}) is equal to the standard
$\text{CP}_s$ transformation~(\ref{eq-simCP}) for
the Higgs fields after a suitable change of
basis, see~(\ref{eq-udef}) and~(\ref{eq-genCPchi}):
\begin{equation}
\varphi'_i(x) \xrightarrow{\CPgm} \varphi'_i{}^{\!*}(x') \quad (i=1,2).
\end{equation}

Using now the results of section~\ref{sec-standCP} we find that
the THDM potential~(\ref{eq-vdef})
will be invariant under a \CPg\ transformation
of type $(ii)$
if and only if we can find a basis
transformation~(\ref{eq-udef}) eliminating
all odd powers of $K_2$. That is,
there must exist some $R(U)\in SO(3)$ such that
\begin{equation}
\label{cptrans}
\begin{split}
\tvec{\xi}' &= R(U)\,\tvec{\xi} = 
\begin{pmatrix}
\cdot\\ 0\\ \cdot
\end{pmatrix},\\
\tvec{\eta}' &= R(U)\,\tvec{\eta} =
\begin{pmatrix}
\cdot\\ 0\\ \cdot
\end{pmatrix},\\
E' &= R(U)\,E\,R^\trans(U) =
\begin{pmatrix}
\cdot & 0 & \cdot\\
0 & \cdot & 0\\
\cdot & 0 & \cdot
\end{pmatrix},
\end{split}
\end{equation}
where the dots represent arbitrary entries. Note that the central entry
of $E'$, that is $E'_{22}$, need not vanish, since it corresponds to a quadratic term in $K_2$.
Obviously, the first two conditions correspond to a rotation
of the vector cross product $\tvec{\xi} \times \tvec{\eta}$ into the $2$-direction
which is always achievable 
by suitable rotations around the $1$- and the $3$-axis. 
It is advantageous to formulate the conditions~(\ref{cptrans}) in a way independent of
the chosen basis, so that no rotations of the original parameters have to be performed.
In the following we shall show that the conditions~(\ref{cptrans}) are equivalent to a simple
set of equations. We formulate this result as a theorem.
\begin{theorem}
\label{theorem-CPgii}
The THDM potential $V$~(\ref{eq-vdef}) is invariant under a 
\CPg\ transformation
(\ref{eq-genCP-Y}), (\ref{eq-genCPKvec})
of type $(ii)$ (see (\ref{eq-genCPsum2}))
if and only if
the following set of equations holds:
\begin{align}
\label{cptransin-a}
(\tvec{\xi} \times \tvec{\eta})^\trans\; E \tvec{\xi} &= 0,\\
\label{cptransin-b}
(\tvec{\xi} \times \tvec{\eta})^\trans\; E \tvec{\eta}  &= 0,\\
\label{cptransin-c}
(\tvec{\xi} \times (E \tvec{\xi}))^\trans\; E^2 \tvec{\xi} &= 0,\\
\label{cptransin-d}
(\tvec{\eta} \times (E \tvec{\eta}))^\trans\; E^2 \tvec{\eta} &= 0.
\end{align}
\end{theorem}
The conditions~(\ref{cptransin-c}) and (\ref{cptransin-d}) are required 
for the case $\tvec{\xi} \times \tvec{\eta} =0$, 
which leads to trivial equations for~(\ref{cptransin-a}) and (\ref{cptransin-b}) and
thus gives no constraints on the matrix~$E$.
By insertion of the explicit expressions~(\ref{cptrans}) 
it is seen that they are sufficient to satisfy~(\ref{cptransin-a})-(\ref{cptransin-d}).
The proof that~(\ref{cptransin-a})-(\ref{cptransin-d}) are
also necessary conditions for~(\ref{cptrans}) to hold
is more lengthy and thus is postponed to the appendix~\ref{sec-independent}.
Since (\ref{cptransin-a})-(\ref{cptransin-d}) just express linear dependencies
of three-vector type quantities via vanishing triple products,
it is obvious that these conditions are rotationally invariant.
They are therefore independent of
the chosen basis,
that is independent of 
transformations~(\ref{eq-partrafo}) of the parameters.
Thus we have found very simple and basis 
independent conditions~(\ref{cptransin-a})-(\ref{cptransin-d}) which are satisfied if and only if
the THDM Higgs potential allows for a \CPg\ symmetry of type~$(ii)$.

The conditions~(\ref{cptransin-a})-(\ref{cptransin-d})
are equivalent to (23)-(26) in~\cite{Gunion:2005ja}
as well as to the conditions given in (A)-(B) in \cite{Nishi:2006tg}.
The proof in appendix~\ref{sec-independent} shows how a Higgs basis is constructed
for which the potential is invariant under the standard \CP\ transformation,
provided (\ref{cptransin-a})-(\ref{cptransin-d}) hold.
In this basis the parameters of the potential with respect to the Higgs fields,
$V(\varphi_1,\varphi_2)$, are real.
Note that by construction the parameters of $V(\fvec{K})$ are always real,
independent of its \CP\ properties.

We remark that the conditions~(\ref{cptransin-a})-(\ref{cptransin-d})
guarantee that the potential has at least one
\CPg\ invariance transformation. It is possible that
a theory has more than one \CPg\ invariance transformation.
A sufficient condition guaranteeing the uniqueness
of the \CPg\ transformation is
\begin{equation}
\tvec{\xi} \times \tvec{\eta} \neq 0.
\end{equation}
Then, clearly the only reflection symmetry one can 
have is on the plane spanned by $\tvec{\xi}$ 
and $\tvec{\eta}$. In appendix~\ref{sec-independent-spon},
table~\ref{t-typeiisym}, we give a
classification of \CPg\ type~$(ii)$
invariant theories with respect to the number of 
independent \CPg\ transformations they allow.

An additional
remark concerns the relation of type~$(i)$
and~$(ii)$ symmetries.
From theorems~\ref{theorem-CPgi} and \ref{theorem-CPgii}
we see that a theory having the \CPg\ symmetry of type $(i)$
is also invariant under - in fact, several -
\CPg\ transformations of type~$(ii)$. 
This is further
discussed in appendix~\ref{sec-independent-spon}.

Eventually we note, that we have classified the \CPg\
properties of the THDM according to the Higgs potential,
regardless of whether these symmetries are spontaneously broken
or not. 
Such a classification of symmetries at the Lagrangian level
is interesting by itself for several reasons: through 
symmetries the parameters of the theory can be restricted.
Moreover, at high temperature one expects to see the full symmetries
of the theory explicitly. In particular, the phase structure
of the theory will depend crucially on these symmetries. Symmetries
may also point the road to generalisations of the theory
relevant at higher energy scales.

In the following section we study in detail 
the conditions for spontaneous breaking of these \CPg\
symmetries.


\section{Spontaneous \CP\ violation}
\label{sec-sponCP}

If there is no \CP\ transformation under which the potential is invariant,
\CP\ is broken \emph{explicitly}.
If the potential is invariant under a certain \CPg\ transformation
but the vacuum expectation value does not respect this symmetry
we have \emph{spontaneous} violation of this \CPg\ symmetry.
Note that a potential can be symmetric under several \CPg\ 
transformations where some may be conserved and some violated
by the vacuum expectation value. Examples for this case 
are given below.

The stationary points of $V$~(\ref{eq-vdef}) with
the lowest potential value give the vacuum solutions for 
$\fvec{K}(x)$~(\ref{eq-Kfour}) and for
the fields. We denote the corresponding
values by
\begin{equation}
\label{eq-vacparam}
\langle \varphi_i \rangle :=
\langle \varphi_i(x) \rangle =
\begin{pmatrix}
v_i^+\\
v_i^0
\end{pmatrix}
\end{equation}
with $i=1,2$.
We get then for the vacuum expectation
values of the matrices $\phi$~(\ref{eq-genCPphi})
and $\twomat{K}$~(\ref{eq-kmat}):
\begin{align}
\label{eq-phivac}
&\langle \phi \rangle :=
\langle \phi(x) \rangle =
\begin{pmatrix}
v_1^+ & v_1^0\\
v_2^+ & v_2^0
\end{pmatrix},\\
\label{eq-kmatvac}
&\twomat{K} =
\frac{1}{2} \big(
K_0 \unitmatrix_2 + \tvec{K} \tvec{\sigma} \big)=
\langle \phi \rangle \langle \phi \rangle^\dagger.
\end{align}
Note that the gauge invariant
functions are
written with argument in this section as
$K_0(x)$, $\tvec{K}(x)$, whereas
the vacuum expectation values
are written without argument,
$K_0$, $\tvec{K}$.
Of course, for an acceptable theory the physical
vacuum must accomplish electroweak symmetry
breaking~(EWSB). That is, the gauge 
group \eweakgroup\ must be broken down
to \emgroup.
In~\cite{Maniatis:2006fs} it has been shown
that this requires
\begin{equation}
\label{eq-K0cond}
K_0 = |\tvec{K}| >0.
\end{equation}
That is, the vacuum solution for the Higgs fields must
correspond to a non-zero light-like four-vector~$\fvec{K}$.
This four-vector~$\fvec{K}$ satisfies the stationarity condition
(see (96) and (145) of~\cite{Maniatis:2006fs} 
and~(\ref{eq-Vfour}) and (\ref{eq-fourpar}))
\begin{equation}
\label{eq-stationarityzero4}
\fvec{\xi}= -2 \left( \fmat{E} - u_0 \fmat{g} \right) \fvec{K}, 
\quad
u_0= \frac{m_{H^\pm}^2}{2v_0^2},
\end{equation}
where
\begin{equation}
\label{eq-gmat}
\fmat{g} := \diag(1,-1,-1,-1),
\end{equation}
or written out in components
\begin{align}
\label{eq-stationarityzero}
\xi_0 &= -2 \big(\eta_{00} K_0 -\frac{m_{H^\pm}^2}{2v_0^2}K_0 +\tvec{\eta}^\trans\tvec{K}\big),\\
\label{eq-stationaritytvec}
\tvec{\xi}&= -2 \big(E \tvec{K}+\frac{m_{H^\pm}^2}{2v_0^2}\tvec{K}+K_0\tvec{\eta}\big).
\end{align}
Here $m_{H^\pm}$ is the mass of the charged Higgs bosons
and
\begin{equation}
v_0 \approx 246\; \text{GeV}
\end{equation}
is the standard Higgs vacuum expectation value.

Suppose now that the potential $V$ has a
\CPg\ symmetry, that is, an
invariance under an improper
rotation $\bar{R}_\varphi$.
The potential parameters satisfy
then~(\ref{eq-genCP1}). This
symmetry is spontaneously broken
if and only if the vacuum
expectation value $\tvec{K}$
does not respect this symmetry,
that is, fulfills
\begin{equation}
\label{eq-sponCP}
\bar{R}_\varphi \tvec{K} \neq \tvec{K}.
\end{equation}
Note the gauge invariance and basis
independence of this condition.

We shall now study the \CPg\
transformations of the
cases~$(i)$ and $(ii)$
separately and discuss
then the standard transformation~$\text{CP}_s$.

\subsection{\CPg\ invariance of type~$(i)$}
\label{sec-subspi}

According to theorem~\ref{theorem-CPgi} the
potential having \CPg\ invariance
of type~$(i)$ has the form
(see~(\ref{eq-genCPtheorem}))
\begin{equation}
\label{eq-sponVi}
V = \xi_0 K_0(x) + \eta_{00} K_0(x)^2 + \tvec{K}(x)^\trans E \tvec{K}(x).
\end{equation}
From~(\ref{eq-K0cond}) we see that 
the correct EWSB requires \mbox{$\tvec{K}\neq 0$}.
This implies then~(\ref{eq-sponCP}) with
$\bar{R}_\varphi=-\unitmatrix_3$. That is, we have 
\begin{equation}
-\unitmatrix_3 \tvec{K} \neq \tvec{K}.
\end{equation}

We formulate this result as a theorem:
\begin{theorem}
\label{theorem-sponCPEWSB}
A theory which is invariant under the
\CPg\ type $(i)$ transformation has the
potential~(\ref{eq-sponVi}).
The required EWSB implies that the
\CPg\ type $(i)$ symmetry is
spontaneously broken.
\end{theorem}
In appendix~\ref{sec-ap-stability}
we discuss in detail the stability and EWSB properties of
this class of models having the potential~(\ref{eq-sponVi}).
There we prove the following theorem.
\begin{theorem}
\label{theorem-sponCP2}
Consider the Higgs part of the 
THDM Lagrangian~(\ref{eq-HiggsL}) with
the potential~(\ref{eq-sponVi}) having
\CPg\ invariance of type~($i$).
Let $\mu_1 \ge \mu_2 \ge \mu_3$ be
the eigenvalues of $E$ with this ordering.
The theory is stable, has the correct EWSB
and no zero mass charged Higgs boson if 
and only if
\begin{equation}
\label{eq-spcptypei}
\begin{split}
&\eta_{00}>0,\\
&\mu_a+ \eta_{00}>0 \quad \text{for}\; a=1,2,3 ,\\
&\xi_0<0,\\
&\mu_3<0.
\end{split}
\end{equation}
The \CPg\ symmetry of type~($i$)
is then spontaneously broken.
\end{theorem}
This clarifies the case of THDM models with type~$(i)$ \CPg\
symmetry completely.


\subsection{\CPg\ invariance of type~$(ii)$}
\label{sec-subspii}

For a theory having a
\CPg\ invariance of type $(ii)$
the parameters of the potential $V$ must
satisfy~(\ref{cptransin-a})-(\ref{cptransin-d})
according to theorem~\ref{theorem-CPgii}.
Such a \CPg\ symmetry is spontaneously
broken if~(\ref{eq-sponCP}) holds
with~$\bar{R}_\varphi$ as in~(\ref{eq-genCPsum2}).
Suppose now that for given parameters 
satisfying~(\ref{cptransin-a})-(\ref{cptransin-d})
it has been checked that~$V$ is a stable potential.
Suppose furthermore, that the vacuum solution~$\twomat{K}$
(\ref{eq-kmatvac}) has been identified. For this we can use,
for instance, the methods of~\cite{Maniatis:2006fs}.
The following theorem allows us then to check if \CPg\ is
spontaneously violated or not.
\begin{theorem}
\label{theorem-spCPgii}
Suppose that the potential is invariant under one or more \CPg\ type (ii) transformations,
that is, its parameters respect~(\ref{cptransin-a})-(\ref{cptransin-d}).
Let~$K_0, \tvec{K}$ be the vacuum solution.
The question if there is a \CPg\ invariance which is also
respected by the vacuum can be decided by checking
the following three relations:
\begin{align}
\label{eq-spcpi1}
( \tvec{\xi} \times \tvec{\eta})^\trans  \tvec{K}  &= 0,\\
\label{eq-spcpi2}
(\tvec{\xi} \times (E \tvec{\xi}))^\trans \tvec{K}  &= 0,\\
\label{eq-spcpi3}
(\tvec{\eta} \times (E \tvec{\eta}))^\trans \tvec{K}  &= 0.
\end{align}
We distinguish two cases.
\begin{equation*}
\label{eq-cross1} 
(a) \quad \tvec{\xi} \times \tvec{\eta} \neq 0.
\end{equation*}
The theory allows then exactly for
one \CPg\ type $(ii)$ invariance transformation which is
conserved also by the vacuum if and
only if~(\ref{eq-spcpi1}) holds. 
In this case~(\ref{eq-spcpi2}) and~(\ref{eq-spcpi3}) are
a consequence of~(\ref{eq-spcpi1}).

\begin{equation*}
\label{eq-cross2} 
(b) \quad \tvec{\xi} \times \tvec{\eta} = 0.
\end{equation*}
Then~(\ref{eq-spcpi1}) is trivial.
There may be more than one \CPg\ type~$(ii)$
invariance transformation. At least one of
these symmetries is also respected by the vacuum
if~(\ref{eq-spcpi2}) and~(\ref{eq-spcpi3}) hold.
\end{theorem}
The proof of theorem~\ref{theorem-spCPgii} is
presented in appendix~\ref{sec-independent-spon}.
We find that the conditions~(\ref{eq-spcpi1})-(\ref{eq-spcpi3})
for the absence of spontaneous \CP\ violation
are equivalent to the conditions given in theorem~4 of \cite{Gunion:2005ja}, 
which were proven in
\cite{Davidson:2005cw} and found before in~\cite{Lavoura:1994fv, Botella:1994cs}.
We find that the criteria a)-c) in~\cite{Nishi:2006tg} correspond to
(\ref{eq-spcpi1})-(\ref{eq-spcpi2}) and should be supplemented by 
(\ref{eq-spcpi3}) to cover the fully general case.
We give the details in appendix~\ref{sec-independent-spon}.

We emphasise that the formulation {\em absence of spontaneous
\CP\ violation} is not quite appropriate in this context.
The correct statement is given in theorem~\ref{theorem-spCPgii} above.
It covers also the case that the theory has more than one 
independent \CPg\ type~$(ii)$ invariance transformation where
one is respected by the vacuum and another spontaneously broken.
These {\em mixed} cases in fact occur; see 
appendix~\ref{sec-independent-spon}.

As discussed in the previous subsection, a type $(i)$ symmetry
is necessarily spontaneously broken in an acceptable theory.
On the other hand, a type $(i)$ symmetric model has at least three
type $(ii)$ symmetries.
It is straightforward to verify that the vacuum respects at least one
of these symmetries; see appendix~\ref{sec-independent-spon}.

To check the 
conditions~(\ref{eq-spcpi1})-(\ref{eq-spcpi3})
we have to know the vacuum expectation value~$\fvec{K}$.
In theorem~2 of~\cite{Maniatis:2006fs}
a classification of all stationary solutions as type~(Ia) to (III) has been given,
covering in particular the vacuum solution.
We discuss in appendix~\ref{sec-ap-stability-ii} 
two necessary conditions for the occurrence of 
spontaneous breaking of a \CPg\ type~$(ii)$
invariance. We formulate this as a theorem.
\begin{theorem}
\label{theorem-sponCPvac}
Spontaneous breaking of a \CPg\
type~$(ii)$ invariance can only
occur if the vacuum solution is of 
type (IIb) (see theorem~2 of~\cite{Maniatis:2006fs}).
That is, the vacuum value~$\fvec{K}$
must be a solution of~(\ref{eq-stationarityzero4})
where
\begin{equation}
\label{eq-sponiivac}
\det \left(\fmat{E}-u_0 \fmat{g} \right)=0.
\end{equation} 
Furthermore, in the basis~(\ref{cptrans})
we must have
\begin{equation}
\label{eq-sponeta22}
\eta_{22}'=-u_0=-\frac{m_{H^\pm}^2}{2 v_0^2}<0
\end{equation}
if the \CPg\ symmetry, corresponding
to the reflection $R_2$~(\ref{eq-R2}) in
this basis, is spontaneously broken.
\end{theorem}


\subsection{$\text{CP}_s$ invariance}
\label{sec-sponcps}

This is, of course, a special case of \CPg\ invariance
of type~($ii$). But now it is convenient to discuss
the situation with respect to the distinguished basis
where the \CP\ transformation
is of the standard type
(see~(\ref{eq-simCP-K}), (\ref{eq-rotK})),
\begin{equation}
\label{eq-CPgii}
\begin{split}
K_0(x)
 &\xrightarrow{\CPs}
K_0(x'),\\
\tvec{K}(x)\;
 &\xrightarrow{\CPs}
R_2 \tvec{K}(x'),
\end{split}
\end{equation}
with $R_2$ the reflection on the $1$--$3$~plane,
see~(\ref{eq-R2}).
Spontaneous $\text{CP}_s$ violation
means in this basis,
from~(\ref{eq-sponCP}) with $\bar{R}_\varphi=R_2$,
that the vacuum does not respect this symmetry:
\begin{equation}
\label{eq-sponCPbasis}
R_2 \tvec{K} \neq \tvec{K},
\end{equation}
that is, we have
\begin{equation}
\label{eq-spcp}
K_2 \neq 0.
\end{equation}

An acceptable theory must have 
a physical vacuum which breaks
\eweakgroup\ down to \emgroup. In this case
the vacuum expectation values of the Higgs doublets may 
be parametrised by
\begin{equation}
\label{eq-vev}
\langle \varphi_1 \rangle =
\begin{pmatrix}
0\\ v_1
\end{pmatrix},
\qquad
\langle \varphi_2 \rangle =
\begin{pmatrix}
0\\ v_2 \;e^{i \zeta}
\end{pmatrix}.
\end{equation}
Here $v_1,v_2,\zeta$ are real numbers with $v_1\ge 0$, $v_2\ge 0$,
\mbox{$-\pi < \zeta \le \pi$},
and a possible phase of $\langle \varphi_1 \rangle$ has been eliminated by a $U(1)_Y$ 
gauge transformation.
The standard Higgs vacuum expectation value is
\begin{equation}
\label{eq-vevex}
v_0 = \sqrt{2(v_1^2+v_2^2)} \approx 246\; \text{GeV}.
\end{equation}
For $v_1 \neq 0$ the usual mixing parameter~$\tan \beta$ can be defined as
$\tan \beta := v_2/v_1$
with $0 \le \beta<\pi/2$.
The vacuum expectation values of the
gauge invariant functions are determined from~(\ref{eq-vacparam})-(\ref{eq-kmatvac})
with~(\ref{eq-vev}) as
\begin{equation}
\label{eq-unit}
\fvec{K} =
\begin{pmatrix}
v_1^2 + v_2^2\\
2 v_1 v_2 \cos \zeta\\
2 v_1 v_2 \sin \zeta\\
v_1^2-v_2^2
\end{pmatrix}.
\end{equation}
From~(\ref{eq-spcp}) we find the
well known result that~$\text{CP}_s$ is 
violated spontaneously if and only 
if~$v_1 \ne 0$, $v_2 \ne 0$, $\zeta \neq 0$ or $\pi$.
That is, the vacuum expectation values of the
two Higgs fields in this special basis must
be complex relative to each other.
We note, however, that this
statement has no basis-independent
meaning. 
Concerning a detailed discussion of
this point see also~\cite{Haber:2006ue}.
By a suitable basis transformation
we can always achieve that only one Higgs
doublet has a non-vanishing vacuum
expectation value which, moreover, is real.
See chapter~6 of~\cite{Maniatis:2006fs}.

At the end of this chapter we make some general
remarks concerning the 
parameters of the THDM 
potential~(see (\ref{eq-vdef}), (\ref{eq-Vfour})).
From~(\ref{eq-stationarityzero4}) it looks
tempting to replace~$\fvec{\xi}$ by the
stationarity condition with~$\fvec{K}$ given in~(\ref{eq-unit})
 (and $v_1$ eliminated by means of (\ref{eq-vevex})):
\begin{equation}
\label{eq-vacuumparam}
\fvec{\xi} = \fvec{\xi}(v_0, v_2,\zeta, m_{H^\pm}, \eta_{00}, \tvec{\eta}, E).
\end{equation}
With this
the potential can be reparametrised in terms of 
$v_0, v_2,\zeta, m_{H^\pm}, \eta_{00}, \tvec{\eta}, E$.
With this set of independent input parameters,
$v_0$ can be adjusted to the required value~(\ref{eq-vevex}),
and relations involving the vacuum solution, such as
the \CP\ invariance conditions~(\ref{cptransin-a})-(\ref{cptransin-d}),
can be evaluated directly in terms of input parameters.
Note, that this parametrisation~(\ref{eq-vacuumparam})
is possible for all potentials having a non-zero stationary point
$\fvec{K}$ on the light cone.
A potential not having such a point can not have the required EWSB behaviour. 
After the substitution~(\ref{eq-vacuumparam})
the four-vector~$\fvec{K}$ in~(\ref{eq-unit}) 
corresponds
by construction
to a stationary point of~$V$. 
Thus, the parametrisation~(\ref{eq-vacuumparam}) is
possible for all potentials with a stationary
point at the wanted place~(\ref{eq-unit}).
But for any concrete values of the new parameters
it remains to be checked whether~$\fvec{K}$ 
in~(\ref{eq-unit})
 is indeed the global minimum
of a stable potential~$V$. This
typically requires to make the complete
analysis of stability and EWSB for~$V$,
for instance with the methods
of~\cite{Maniatis:2006fs}.
Note that in the gauge invariant function approach this change of
parameters is even possible for the cases where the phase
$\zeta$ or one of $v_1, v_2$ vanishes.


\section{Examples}
\label{sec-exam}

Here we apply the general considerations of Sections~\ref{sec-explCP}
and~\ref{sec-sponCP} to specific models.


\subsection{\CP\ symmetric model with $\tvec{\xi}=\tvec{\eta}=0$ }
\label{ssec-thirdex}

We consider the~THDM with the Higgs potential
\begin{equation}
\begin{split}
\label{eq-newp}
V(\varphi_1,\varphi_2)&=
m_{11}^2 \left(\varphi_1^\dagger \varphi_1
+ \varphi_2^\dagger \varphi_2 \right)\\
&
+\frac{1}{2} \lambda_1 \left((\varphi_1^\dagger \varphi_1)^2 + (\varphi_2^\dagger \varphi_2)^2 \right)\\
&
+\lambda_3 (\varphi_1^\dagger \varphi_1) (\varphi_2^\dagger \varphi_2)
+\lambda_4 (\varphi_1^\dagger \varphi_2) (\varphi_2^\dagger \varphi_1)\\
&
+\frac{1}{2} \lambda_5 \left( (\varphi_1^\dagger \varphi_2)^2 + (\varphi_2^\dagger \varphi_1)^2 \right),
\end{split}
\end{equation}
where all parameters are real.
This potential is invariant under
$\varphi_1 \longrightarrow -\varphi_1$.
We put the
potential into the form~(\ref{eq-vdef}) using the
relations~(\ref{eq-phik}).  Then,
\begin{equation}
\begin{split}
\label{eq-pai}
\eta_{00} &= \frac{1}{4}(\lambda_1+\lambda_3),\\
\tvec{\eta}
  &= \begin{pmatrix} 0\\ 0\\ 0 \end{pmatrix},\\
E &= \frac{1}{4} \begin{pmatrix}
    \lambda_4 + \lambda_5 & 0 & 0\\
    0 & \lambda_4 - \lambda_5 & 0\\
    0 & 0 & \lambda_1 - \lambda_3 \end{pmatrix},\\
\xi_0 &= m_{11}^2,\\
\tvec{\xi}
  &= \begin{pmatrix} 0\\ 0\\ 0 \end{pmatrix}.
\end{split}
\end{equation}
Obviously this model fulfills the conditions of 
theorem~\ref{theorem-CPgi}, that is, has a \CPg\ symmetry
of type~$(i)$.
Furthermore, the potential has at least three \CPg\ symmetries of
type~$(ii)$, namely $R_1,R_2,R_3$,
and infinitely many if two or three eigenvalues of $E$ coincide.
Note that the condition 
$\tvec{\xi}=\tvec{\eta}=0$ is basis independent. 
This in turn means that every potential with
$\tvec{\xi}=\tvec{\eta}=0$ can be cast
into the form~(\ref{eq-newp}) respectively (\ref{eq-pai}) with an appropriate
basis transformation. 

We see from theorem~\ref{theorem-sponCP2}
and the discussion in appendix~\ref{sec-ap-stability}
that this model is stable {\em in the strong sense}
if simultaneously
$\lambda_1>0$,
$\lambda_1+\lambda_3>0$ and
$\lambda_1+\lambda_3+\lambda_4>|\lambda_5|$.
Moreover, it
has the right electroweak symmetry breaking
behaviour for
$\xi_0<0$ or equivalently $m_{11}^2 < 0$.
In the case of~$m_{11}^2 < 0$
the \CPg\ symmetry of type~$(i)$ is spontaneously broken.
However, at least one \CPg\ symmetry of type~$(ii)$ is respected
by the vacuum; see also appendix~\ref{sec-independent-spon}.


\subsection{\CP\ properties of the ``almost general'' THDM}
\label{ssec-firstex}

We consider a class of THDMs with the Higgs potential
\begin{equation}
\begin{split}
\label{eq-gunp}
V(\varphi_1,\varphi_2)&=
m_{11}^2 \varphi_1^\dagger \varphi_1
+ m_{22}^2 \varphi_2^\dagger \varphi_2\\
&
-\bigg[m_{12}^2 \varphi_1^\dagger \varphi_2 + h.c.\bigg]\\
&
+\frac{1}{2} \lambda_1 (\varphi_1^\dagger \varphi_1)^2
+\frac{1}{2} \lambda_2 (\varphi_2^\dagger \varphi_2)^2\\
&
+\lambda_3 (\varphi_1^\dagger \varphi_1) (\varphi_2^\dagger \varphi_2)
+\lambda_4 (\varphi_1^\dagger \varphi_2) (\varphi_2^\dagger \varphi_1)\\
&
+\bigg[
        \frac{1}{2} \lambda_5 (\varphi_1^\dagger \varphi_2)^2 + h.c.
\bigg],
\end{split}
\end{equation}
written in the parametrisation of~\cite{Haber:1993an}, where
 $m_{12}^2$ and $\lambda_5$ may be arbitrary complex and
all other parameters are real.
This potential breaks the discrete symmetry
$\varphi_1 \longrightarrow -\varphi_1$
only softly, that is by quadratic terms in the Higgs doublet fields,
thus suppressing large flavour-changing neutral currents. 
We put the
potential into the form~(\ref{eq-vdef}) using the
relations~(\ref{eq-phik}) and get here
\begin{equation}
\begin{split}
\label{eq-pagu}
\eta_{00}
  &= \frac{1}{8}(\lambda_1 + \lambda_2 + 2 \lambda_3),\\[.2cm]
\tvec{\eta}
  &= \frac{1}{8} \begin{pmatrix} 0\\ 0\\ \lambda_1 - \lambda_2 \end{pmatrix},
 \\[.2cm]
E &= \frac{1}{4} \begin{pmatrix}
    \lambda_4+\mRe \lambda_5 & -\mIm \lambda_5 & 0\\
    -\mIm \lambda_5 & \lambda_4- \mRe \lambda_5 & 0\\
    0 & 0 & \frac{1}{2}(\lambda_1+\lambda_2-2 \lambda_3) \end{pmatrix},\\
\xi_0 &= \frac{1}{2}(m_{11}^2+m_{22}^2),\\
\tvec{\xi} &= \begin{pmatrix}
    - \mRe m_{12}^2\\[2mm]
    \phantom{-} \mIm m_{12}^2\\[2mm]
    \frac{1}{2} \left( m_{11}^2-m_{22}^2 \right)\\[2mm]
  \end{pmatrix}.
\end{split}
\end{equation}
The stability of the potential is easily
investigated using the methods of~\cite{Maniatis:2006fs}.
Stability is
guaranteed by the terms
quartic in the fields alone if and only if
\begin{equation}
\lambda_1 > 0, \quad \lambda_2 > 0, \quad\text{and~}
 \sqrt{\lambda_1 \lambda_2} + \lambda_3 > \max(0, \abs{\lambda_5}-\lambda_4).
\end{equation}
In order to determine the \CP\ properties of the potential
we have to check~(\ref{cptransin-a})-(\ref{cptransin-d}).
Two of the conditions for \CPg\ type~$(ii)$ invariance of
the potential, (\ref{cptransin-b})
and~(\ref{cptransin-d}), are, with~(\ref{eq-pagu}), 
automatically fulfilled. 
The remaining conditions~(\ref{cptransin-a}) 
and~(\ref{cptransin-c}) give
\begin{align}
\label{eq-gunionc1}
(\lambda_1 - \lambda_2)\,\mIm\left((m_{12}^2)^2\lambda_5^*\right)&=0\,,\\[2mm]
\label{eq-gunionc2}
\left[\left(\lambda_1+\lambda_2-2(\lambda_3+\lambda_4)\right)^2-4 \abs{\lambda_5}^2
\right] & \nonumber\\
\times (m_{11}^2-m_{22}^2)\,
\mIm\left((m_{12}^2)^2\lambda_5^*\right)&=0
\end{align}
as necessary and sufficient conditions for 
the existence of a \CPg\ invariance
of type~$(ii)$ for the potential.
It is obvious that for
the case 
of real parameters $m_{12}^2$ and $\lambda_5$
(\ref{eq-gunionc1}) and (\ref{eq-gunionc2}) are satisfied.
For $\tvec{\xi} \times \tvec{\eta}$ we find 
from~(\ref{eq-pagu})
\begin{equation}
\tvec{\xi} \times \tvec{\eta} =
\frac{1}{8} (\lambda_1-\lambda_2)
\begin{pmatrix}
\mIm(m_{12}^2)\\
\mRe(m_{12}^2)\\
0
\end{pmatrix}.
\end{equation}
From theorem~\ref{theorem-CPgii} ff. we find, therefore,
that in this model the potential allows one or more
\CPg\ symmetries if and only if~(\ref{eq-gunionc1})
and~(\ref{eq-gunionc2}) hold. There is exactly
one \CPg\ symmetry if \mbox{$\lambda_1-\lambda_2 \neq 0$}
and \mbox{$m_{12}^2 \neq 0$}.

In the case \CPg\ is conserved, that is~(\ref{eq-gunionc1}), (\ref{eq-gunionc2}) 
are fulfilled,
\CPg\ may be violated spontaneously.
We reparametrise the potential using the stationarity conditions
\eqref{eq-stationarityzero}, \eqref{eq-stationaritytvec} and
assume that the vacuum expectation values $v_1$, $v_2$ together with the phase $\zeta$
indeed describe the global minimum~(\ref{eq-unit}) of the potential.
We check the conditions for spontaneous \CPg\ violation \eqref{eq-spcpi1}-\eqref{eq-spcpi3}
and see that \eqref{eq-spcpi3} is automatically fulfilled.
We find that \eqref{eq-spcpi1} and \eqref{eq-spcpi2} together with
\eqref{eq-gunionc1} and \eqref{eq-gunionc2}
are equivalent to the condition that either
\begin{equation}
\label{eq-gunionfullcp1}
v_1 v_2 \left[\cos(2\zeta)\mIm\lambda_5 + \sin(2\zeta)\mRe\lambda_5\right] =0\\
\end{equation}
or
\begin{equation}
\label{eq-gunionfullcp2}
\lambda_1=\lambda_2,
\qquad
(v_1^2-v_2^2)\left[(\lambda_3+\lambda_4-\lambda_1)^2-\abs{\lambda_5}^2\right]=0
\end{equation}
or both are fulfilled.
That is, exactly if \eqref{eq-gunionfullcp1}
or \eqref{eq-gunionfullcp2} or both are fulfilled,
there is a \CPg\ symmetry of both the potential 
and the vacuum expectation value~$\fvec{K}$.


\section{Conclusions}
\label{sec-conclu}
In this work we have shown that
the framework of gauge invariant functions
is well suited to
discuss \CP\ properties of the general THDM. 
These real gauge invariant functions build a four-vector
for which we could reveal
a simple geometric picture: Mixing of the two Higgs doublets corresponds
to rotations 
and \CP\ transformations to reflections of the 
{\em space-like} components of this four-vector.

In this geometric picture we have
first given a classification of possible
\CP\ transformations in the THDM;
see section~\ref{sec-explCP}.
The standard \CP\ transformation
involves no mixing of
the two doublet fields and corresponds to a
reflection on the $1$--$3$ plane. 
We identified
two types, $(i)$ and~$(ii)$, of
generalised \CP\ transformations where
arbitrary unitary mixing of the two
doublet fields is allowed.
The type~$(i)$ \CPg\ transformation is represented by a point reflection
and has, to our knowledge, not been discussed before.
We gave conditions for a theory to be symmetric under this transformation
in theorem~\ref{theorem-CPgi}.
Type~$(ii)$ \CPg\ transformations correspond to reflections on planes
and include in particular the standard \CP\ transformation.
In theorem~\ref{theorem-CPgii} we gave
simple and easy to check conditions
the parameters of the THDM potential
have to satisfy if the Higgs Lagrangian
is to be invariant under a \CPg\ transformation.
We also gave a classification showing which
THDMs allow for just one \CPg\ symmetry and 
which for more than one; see 
table~\ref{t-typeiisym} in 
appendix~\ref{sec-independent-spon}.

In section~\ref{sec-sponCP} we turned
to the question whether in a \CP\
symmetric theory the vacuum respects
the symmetry or not.
We derived necessary and sufficient
conditions for this to be the case.
Again, the conditions to be checked
are in all cases simple and have a
transparent geometric meaning.
See theorems~\ref{theorem-sponCPEWSB}-\ref{theorem-sponCPvac} 
and (\ref{eq-sponCPbasis}),
(\ref{eq-spcp}).
We emphasise that a THDM can
have more than one \CPg\ invariance
where one may be spontaneously broken
while another is conserved by the vacuum.
This is again a crucial element for the discussions
in the companion paper \cite{Maniatis:2007de}.

In section~\ref{sec-exam}
we have illustrated our geometric methods with two examples, 
namely the most general model respecting the \CPg\
type $(i)$ symmetry,
as well as 
a model where large flavour-changing neutral currents are suppressed.

We compared our results with the literature for cases where this is possible.
Concerning the existence of a \CP\ type~(ii) symmetry,
our results fully agree with those of \cite{Davidson:2005cw, Gunion:2005ja},
where a completely different approach was used.
For the existence of a \CP\ type~(ii) symmetry of the potential
we agree with~\cite{Nishi:2006tg}.
But our criteria for \CP\ invariance of the vacuum give
important supplements to those presented in~\cite{Nishi:2006tg}.

As mentioned above we have introduced and discussed in our present paper
as completely new element the \CPg\ symmetry of type~(i).
Furthermore we have given a thorough discussion of the cases where
multiple \CPg\ transformations are allowed by the theory.
All these new elements are essential ingredients for our companion
paper \cite{Maniatis:2007de}.

\begin{acknowledgments}
The authors thank R.~Barbieri for useful discussions.
\end{acknowledgments}

\renewcommand{\theequation}{\Alph{section}.\arabic{equation}}

\appendix

\section{Basis independent conditions for
\CPg~type~$(ii)$~invariance of the potential}
\label{sec-independent}

In this appendix we complete the proof of theorem~\ref{theorem-CPgii}
by showing that the existence of a basis~(\ref{cptrans}),
meaning \CPg\ type $(ii)$ invariance of the potential (see (\ref{eq-genCPsum2})),
is equivalent to (\ref{cptransin-a})-(\ref{cptransin-d}).

We show first that~(\ref{cptrans}) implies~(\ref{cptransin-a})-(\ref{cptransin-d}).
Indeed we have for~$\tvec{\xi}', \tvec{\eta}'$
and $E'$ as in~(\ref{cptrans}) 
\begin{equation}
\begin{alignedat}{2}
\label{eq-pbasis}
\tvec{\xi}' \times \tvec{\eta}' &= 
\begin{pmatrix}
0\\ \cdot \\ 0
\end{pmatrix},\\
E' \tvec{\xi}' &= 
\begin{pmatrix}
\cdot\\ 0 \\ \cdot
\end{pmatrix},\qquad&
E' \tvec{\eta}' &= 
\begin{pmatrix}
\cdot\\ 0 \\ \cdot
\end{pmatrix},\\
\tvec{\xi}' \times E' \tvec{\xi}' &= 
\begin{pmatrix}
0\\ \cdot \\ 0
\end{pmatrix},&
\tvec{\eta}' \times E' \tvec{\eta}' &= 
\begin{pmatrix}
0\\ \cdot \\ 0
\end{pmatrix}.
\end{alignedat}
\end{equation}
Furthermore, for any vector $\tvec{\zeta}_\perp$ with
2-component zero,
\begin{equation}
\tvec{\zeta}_\perp  = 
\begin{pmatrix}
\cdot\\ 0 \\ \cdot
\end{pmatrix},
\end{equation}
we have
\begin{equation}
E' \tvec{\zeta}_\perp  = 
\begin{pmatrix}
\cdot\\ 0 \\ \cdot
\end{pmatrix}.
\end{equation}
Thus, for all vectors $\tvec{\zeta}_\parallel$ of the form
\begin{equation}
\tvec{\zeta}_\parallel  = 
\begin{pmatrix}
0\\ \cdot \\ 0
\end{pmatrix}
\end{equation}
we have
\begin{equation}
\label{eq-zeta}
\tvec{\zeta}_\parallel^\trans E' \tvec{\zeta}_\perp  =0.
\end{equation}
All expressions~(\ref{cptransin-a})-(\ref{cptransin-d}) are of the 
form~(\ref{eq-zeta})
if~(\ref{cptrans}) holds. The conditions
are formulated in a rotationally invariant form.
Thus, they hold for $\tvec{\xi}$, $\tvec{\eta}$,
$E$ if they hold for
$\tvec{\xi}'$, $\tvec{\eta}'$, $E'$, q.e.d.

Now we want to show that from~(\ref{cptransin-a})-(\ref{cptransin-d})
follows~(\ref{cptrans}) with a suitable rotation $R(U)$.
First we choose a basis where
\begin{equation}
\label{eq-spbasis}
\begin{split}
\tvec{\xi}' &= R(U)\,\tvec{\xi} = 
\begin{pmatrix}
0 \\ 0\\ \xi_3'
\end{pmatrix},\\
\tvec{\eta}' &= R(U)\,\tvec{\eta} =
\begin{pmatrix}
\eta_1' \\ 0\\ \eta_3'
\end{pmatrix}.
\end{split}
\end{equation}
Note that there is always a rotation into this basis possible 
for two vectors.
It remains to be shown that
in addition $\eta_{12}'=\eta_{23}'$=0 can be achieved
if~(\ref{cptransin-a})-(\ref{cptransin-d}) hold.
We remark that~$E$ 
is a symmetric matrix~(see~(\ref{eq-vdef})) and
this property is not altered by a similarity 
transformation~(\ref{eq-partrafo}).
We have to consider different cases depending
on whether the vector cross product
\begin{equation}
\label{eq-cross}
\tvec{\xi}' \times \tvec{\eta}' = 
\begin{pmatrix}
0\\ \xi_3' \eta_1'\\ 0
\end{pmatrix}
\end{equation}
vanishes or not.
Let us first assume that the vector cross product~(\ref{eq-cross})
does not vanish, that is, we have $\xi_3' \eta_1' \neq 0$.
From~(\ref{cptransin-a}) we find now
\begin{equation}
(\tvec{\xi}' \times \tvec{\eta}')^\trans\; E'^\; \tvec{\xi}'   = 
\eta_{23}'\; \xi_3'^2 \eta_1'=0.
\end{equation}
This means that $\eta_{23}' =\eta_{32}'=0$. Then~(\ref{cptransin-b}) gives
\begin{equation}
(\tvec{\xi}' \times \tvec{\eta}')^\trans  E'^\; \tvec{\eta}'  =
\eta_{21}'\; \xi_3' \eta_1'^2=0,
\end{equation}
that is, we have also $\eta_{21}' =\eta_{12}'=0$.
Thus, the explicit form~(\ref{cptrans}) follows 
from~(\ref{cptransin-a})-(\ref{cptransin-d}) for this case.

Now we have to consider also the special case of a vanishing 
vector cross product~(\ref{eq-cross}). 
In this case~(\ref{cptransin-a}) and (\ref{cptransin-b})
are trivially fulfilled and give no constraint for the matrix $E'$.
We shall now use~(\ref{cptransin-c}) and (\ref{cptransin-d})
to prove~(\ref{cptrans}).
If $\tvec{\xi} \times \tvec{\eta} =0$ and $\tvec{\xi}=0$
and $\tvec{\eta} =0$ we can achieve~(\ref{cptrans})
trivially by diagonalising $E$. Thus,
consider the case that $\tvec{\xi} \times \tvec{\eta} =0$
and $\tvec{\xi} \neq 0$.
By an orthogonal transformation we can diagonalise $E$:
\begin{equation}
R(U_1) E R^\trans (U_1)=E'=\diag (\mu_1, \mu_2, \mu_3).
\end{equation}
We get then in this basis already
$\eta_{12}'=\eta_{23}'=0$. Furthermore, we 
have
\begin{equation}
\begin{aligned}
&\tvec{\xi}'=
\begin{pmatrix}
\xi_1'\\ \xi_2'\\ \xi_3'
\end{pmatrix},
\qquad
E' \tvec{\xi}'=
\begin{pmatrix}
\mu_1 \xi_1'\\ \mu_2 \xi_2'\\ \mu_3 \xi_3'
\end{pmatrix},
\\
& \tvec{\xi}' \times E'  \tvec{\xi}' =
\begin{pmatrix}
(\mu_3 - \mu_2) \xi_2' \xi_3'\\ 
(\mu_1 - \mu_3) \xi_3'  \xi_1'\\ 
(\mu_2 - \mu_1)\xi_1'  \xi_2'
\end{pmatrix},
\end{aligned}
\end{equation}
and from~(\ref{cptransin-c}), 
\begin{multline}
\label{eq-appeta}
\left(\tvec{\xi}' \times \left(E' \tvec{\xi}'\right) \right)^\trans
 E'^2 \tvec{\xi}'=\\
(\mu_1 - \mu_2)(\mu_2 - \mu_3)(\mu_3 - \mu_1)
\xi_1' \xi_2' \xi_3' =0.
\end{multline}
If all eigenvalues $\mu_a$ are
different we find from~(\ref{eq-appeta})
that at least one $\xi_a'$ must be zero.
By a change of basis which interchanges the components
we can achieve $\xi_2'=0$ without introducing off-diagonal
elements in $E'$.
Then $\tvec{\eta}'$ being parallel to
$\tvec{\xi}'$ implies $\eta_2'=0$
and we found a basis of the form~(\ref{cptrans}).
Suppose, on the other hand, that
at least two eigenvalues $\mu_a$
are equal. Without loss of generality
we can suppose
\begin{equation}
\mu_1=\mu_2.
\end{equation}
By a rotation around the $3$-axis,
leaving $E'$ diagonal,
we can then achieve
\begin{equation}
\tvec{\xi}'=
\begin{pmatrix}
\xi_1'\\ 0\\ \xi_3'
\end{pmatrix}
\end{equation}
and also $\eta_2'=0$ since
$\tvec{\eta}'$ is parallel to $\tvec{\xi}'$, q.e.d.
For the case $\tvec{\xi} \times \tvec{\eta} =0$
and $\tvec{\eta} \neq 0$ the argumentation runs
along the same lines using~(\ref{cptransin-d}) instead
of~(\ref{cptransin-c}). This completes the proof
that the set of the conditions~(\ref{cptransin-a})-(\ref{cptransin-d}) is
equivalent to the existence of a basis satisfying~(\ref{cptrans}).

We compared our conditions~(\ref{cptransin-a})-(\ref{cptransin-d})
for \CP\ invariance of the potential with~(23)-(26) in~\cite{Gunion:2005ja}.
In~\cite{Gunion:2005ja} the conditions were
found by a systematic survey of all possible
complex invariants - and there is an enormous
number of such invariants - within a field based formulation,
that is, in a completely different way.
Our triple products required to vanish in \eqref{cptransin-a},
\eqref{cptransin-b}, and \eqref{cptransin-d} turn out to be equal to
$-2^{-5} I_{2Y2Z}$, $2^{-7} I_{Y3Z}$, and $-2^{-13} I_{6Z}$
in their notation.
Despite the fact, that the fourth invariant occurring in~\cite{Gunion:2005ja} 
and our condition~\eqref{cptransin-c}
are different,
we can show that the full sets of conditions are equivalent.
This is conveniently done by computing the {\em reduced
Groebner bases} for both sets which
are indeed equal (for 
a brief introduction to the
formalism of Groebner bases
see the appendix of~\cite{Maniatis:2006jd}).


\section{Theories with \CPg~type~$(i)$~invariance}
\label{sec-ap-stability}

Here we study the theories having a \CPg\
invariance of type~$(i)$ in detail;
see theorem~\ref{theorem-CPgi}, 
(\ref{eq-genCPtheorem}).
The corresponding potential is given 
in~(\ref{eq-sponVi}).

We show first that the transformation 
(\ref{eq-genCPphieps}) of the fields is unique,
up to gauge transformations, in 
giving the \CPg\ type~$(i)$ transformation
for the gauge invariant functions:
\begin{equation}
\label{B100}
\begin{split}
K_0(x) &\xrightarrow{} K_0(x'),\\
\tvec{K}(x)\; &\xrightarrow{} - \tvec{K}(x') .
\end{split}
\end{equation}
To see this we try to generalise~(\ref{eq-genCPphieps}) by
setting
\begin{equation}
\label{B101}
\phi(x) \xrightarrow{\CPg} V \epsilon \phi^*(x')
\end{equation}
with $V \in U(2)$. Every $V \in U(2)$ can be represented as
\begin{equation}
\label{B102}
V = e^{i \gamma} \tilde{V}
\end{equation}
with $\gamma$ real and $\tilde{V} \in SU(2)$.
The transformation of $\twomat{K}(x)$ and
$K_0(x)$, $\tvec{K}(x)$ induced by (\ref{B101})
reads (see (\ref{eq-genCP-K}) and (\ref{eq-genCP-Kmateps}))
\begin{equation}
\begin{split}
\twomat{K}(x) \xrightarrow{\CPg}& {V} \epsilon\; \twomat{K}^\trans(x')
\epsilon^\trans  V^\dagger\\
=&\tilde{V} \epsilon \twomat{K}^\trans(x')
\epsilon^\trans \tilde{V}^{\dagger}\\
=&\tilde{V}  
 \frac{1}{2} \left(K_0(x') \unitmatrix_2-\tvec{K}(x') \tvec{\sigma}  \right)
\tilde{V}^{\dagger},
\end{split}
\end{equation}
\begin{equation}
\label{B104}
\begin{split}
  K_0(x) &\xrightarrow{\CPg} \phantom{+}K_0(x'),\\
  \tvec{K}(x) &\xrightarrow{\CPg} -R(\tilde{V}) \; \tvec{K}(x').
\end{split}
\end{equation}
Here $R(\tilde{V})$ is obtained from~(\ref{eq-defrot})
with $U$ replaced by $\tilde{V}$. In order to obtain the \CPg\ transformation
of type $(i)$ from (\ref{B104}) we must have
\begin{equation}
R(\tilde{V})=\unitmatrix_3.
\end{equation}
which implies
\begin{equation}
\tilde{V}= \pm \unitmatrix_2
\end{equation}
since $\tilde{V} \in SU(2)$. From (\ref{B101}) and (\ref{B102})
we get, therefore, as the only possible transformations of the
fields leading to a \CPg\ transformation of type~$(i)$
\begin{equation}
\label{B107}
\phi(x) \xrightarrow{} e^{i \gamma} \epsilon \; \phi^*(x'),
\end{equation}
where in the case $\tilde{V}=-\unitmatrix_2$ we have redefined 
$\gamma$ as $\gamma+\pi$. Both Higgs doublets have weak
hypercharge $y=+1/2$. 
Thus a gauge transformation $U_G \equiv \exp( -2 i \gamma {\bf Y})$
brings back~(\ref{B107}) to the form~(\ref{eq-genCPphieps})
\begin{equation}
\label{B108}
\phi(x) \xrightarrow{\CPg} e^{i \gamma} \epsilon \; \phi^*(x')
\xrightarrow{U_G} \epsilon \; \phi^*(x'),
\end{equation}
as we asserted.

We note that our arguments are valid in any basis. Thus, the 
transformation~(\ref{eq-genCPphieps}) has the
interesting property of being the same, independently
of the choice of basis. This holds
again up to gauge transformations. We can also
see this directly from 
(\ref{eq-genCPchi})-(\ref{eq-genCPRtildechi}). 
We start from~(\ref{eq-genCPphieps}) and
make a basis transformation (\ref{eq-genCPchi})
with $U \in U(2)$. Then we get from (\ref{eq-genCPdefUchi})
with $U_\varphi=\epsilon$
\begin{equation}
\label{B109}
U'_\varphi = U \epsilon U^{*-1}.
\end{equation}
We can decompose $U$ as
\begin{equation}
U = e^{i \gamma/2} \tilde{U}
\end{equation}
with $\gamma$ real and $\tilde{U}\in SU(2)$. For any
$\tilde{U}\in SU(2)$ we have
\begin{equation}
\epsilon^\trans \tilde{U} \epsilon = \tilde{U}^*.
\end{equation}
Inserting this in (\ref{B109}) we get
\begin{equation}
\begin{split}
U'_\varphi =& e^{i \gamma} \tilde{U} \;\epsilon\; \tilde{U}^{* -1}\\
=& e^{i \gamma}\; \epsilon\; \epsilon^\trans \tilde{U} \epsilon  \tilde{U}^{* -1}\\
=& e^{i \gamma}\; \epsilon\; \tilde{U}^*  \tilde{U}^{* -1}\\
=& e^{i \gamma}\; \epsilon.
\end{split}
\end{equation}
Again, the factor $\exp(i \gamma)$ just represents a gauge transformation. With
this we have shown directly the basis independence of the $\CPg$ transformation
of type~$(i)$ given in~(\ref{eq-genCPphieps}).

We go now to a basis where $E$ is diagonal,
\begin{equation}
\label{eq-sponEdiag}
E=\diag (\mu_1,\mu_2,\mu_3)
\end{equation}
with the ordering 
\begin{equation}
\label{eq-sponEdiagord}
\mu_1 \geq \mu_2 \geq \mu_3.
\end{equation}

For the discussion of the stability
of the theory we have to consider
the function $f(u)$ (see~(55) of~\cite{Maniatis:2006fs})
and the set $I$ of $u$ values
defined in (70) of~\cite{Maniatis:2006fs}.
Here we find
\begin{equation}
\label{eq-sponCPf}
\begin{split}
f(u) &= u + \eta_{00},\\
f'(u) &= 1,
\end{split}
\end{equation}
\begin{equation}
I= \left\{0,\mu_1, \mu_2, \mu_3 \right\}.
\end{equation}
Now we go through the criteria spelled
out in theorems~1-3 in~\cite{Maniatis:2006fs}
which tell us when the theory is stable and has
the correct EWSB behaviour. 
In view of theorem~1 of~\cite{Maniatis:2006fs}
we see that
stability in the strong sense
requires
\begin{equation}
\label{eq-ap-stabstr}
\begin{split}
f(0) &= \eta_{00} >0,\\
f(\mu_a) &= \mu_a + \eta_{00} >0 \quad (a=1,2,3).
\end{split}
\end{equation}

If $\eta_{00}=0$ or $\mu_a+\eta_{00}=0$
for at least one $a \in \{1,2,3\}$
we have to consider the function~$g(u)$,
see (72) of~\cite{Maniatis:2006fs}.
Here we get
\begin{equation}
g(u)=\xi_0.
\end{equation}
Stability in the {\em weak sense} requires
then $\xi_0>0$, {\em marginal}
stability $\xi_0=0$. 
On the other
hand, we have from~(117) of~\cite{Maniatis:2006fs}
the necessary condition for EWSB $\xi_0 < |\tvec{\xi}|$
which gives here 
\begin{equation}
\label{eq-ap-EWSBi}
\xi_0<0.
\end{equation}
Thus we find that a potential~(\ref{eq-sponVi})
being stable in the {\em weak sense}
or only {\em marginally} stable cannot
have the correct EWSB.
In other words: in an acceptable
theory of this kind the
potential parameters must satisfy~(\ref{eq-ap-stabstr})
and (\ref{eq-ap-EWSBi}). This already proves the
first three
relations~(\ref{eq-spcptypei}) of theorem~\ref{theorem-sponCP2}.

Next we study the stationary points
of $V$, (\ref{eq-sponVi}),
using the four-dimensional 
notation~(\ref{eq-Kfour})-(\ref{eq-fourpar}).
The constraints~(\ref{eq-kinq})
on the gauge invariant functions read
\begin{equation}
\label{eq-domv}
\fvec{K}(x)^\trans \fmat{g} \fvec{K}(x) \geq 0,
\qquad K_0(x) \ge 0,
\end{equation}
with~$\fmat{g}$ given in~(\ref{eq-gmat}). 
For the potential~(\ref{eq-sponVi}) we have,
with~(\ref{eq-ap-EWSBi}) and~(\ref{eq-sponEdiag}),
\begin{equation}
\fvec{\xi} =
\begin{pmatrix}
\xi_0\\
0\\
0\\
0
\end{pmatrix},
\quad
\xi_0<0,
\end{equation}
\begin{equation}
\fmat{E}= \diag (\eta_{00}, \mu_1, \mu_2, \mu_3 ).
\end{equation}
The point $\fvec{K}=0$ is always a stationary solution.
We now check for the non-trivial stationary points.
In the interior of the forward
light cone, the stationary points of $V$  are obtained from (91) of~\cite{Maniatis:2006fs},
\begin{align}
\label{eq-ap-interior-1}
&\fmat{E} \fvec{K} = -\frac{1}{2} \fvec{\xi},\\
\label{eq-ap-interior-2}
&\fvec{K}^\trans \fmat{g} \fvec{K} > 0,\\
\label{eq-ap-interior-3}
&K_0 >0.
\end{align}
From~(\ref{eq-ap-interior-1}) we get here
\begin{equation}
\begin{split}
\eta_{00} K_0 &= -\frac{1}{2} \xi_0,\\
\mu_1 K_1 &= 0,\\
\mu_2 K_2 &= 0,\\
\mu_3 K_3 &= 0.
\end{split}
\end{equation}
It follows that
\begin{equation}
K_0=\frac{1}{2 \eta_{00}} \big( -\xi_0 \big) >0.
\end{equation}
Thus~(\ref{eq-ap-interior-3})
is already fulfilled. 
If
\begin{equation}
\det E= \mu_1 \mu_2 \mu_3 \neq 0,
\end{equation}
the only stationary point in the interior of
the light cone is, therefore,
\begin{equation}
\label{eq-regsol}
\fvec{K}=
-\frac{\xi_0}{2 \eta_{00}}
\begin{pmatrix}
1\\
0\\
0\\
0
\end{pmatrix}.
\end{equation}
For $\det E=0$ we have regions of stationary
points extending from the solution~(\ref{eq-regsol})
to the light cone.

A vacuum with the required EWSB must lie on the forward light cone.
We now study all stationary points in this part of the domain,
see (96) of~\cite{Maniatis:2006fs}:
\begin{align}
\label{eq-ap-stat1}
&(\fmat{E}- u \fmat{g}) \fvec{K}= -\frac{1}{2} \fvec{\xi},\\
\label{eq-ap-stat2}
&\fvec{K}^\trans \fmat{g} \fvec{K} = 0,\\
\label{eq-ap-stat3}
&K_0 >0.
\end{align}
From~(\ref{eq-ap-stat1}) we get
\begin{equation}
\label{eq-stat-exp}
\begin{split}
(\eta_{00}- u)  K_0 &= -\frac{1}{2} \xi_0,\\
(\mu_1 + u)  K_1 &= 0,\\
(\mu_2 + u)  K_2 &= 0,\\
(\mu_3 + u)  K_3 &= 0.
\end{split}
\end{equation}
For the functions $\tilde{f}(u)$ and
$\tilde{f}'(u)$, (102) and (103)
of~\cite{Maniatis:2006fs}, we find
\begin{align}
\tilde{f}(u) &= -\frac{1}{4} \frac{\xi_0^2}{\eta_{00}-u},\\ 
\tilde{f}'(u) &= -\frac{1}{4} \frac{\xi_0^2}{(\eta_{00}-u)^2}.
\end{align}

Now we use theorem~2 of~\cite{Maniatis:2006fs}
to discuss the stationary points of~$V$.
Since we have here always
\begin{equation}
\tilde{f}'(u)<0
\end{equation}
there are no solutions of type~$(IIa)$.
But there are solutions of type~$(IIb)$,
that is, solutions with
\begin{equation}
\det (\fmat{E} - u \fmat{g} )=0.
\end{equation}
These occur for
\begin{equation}
u=-\mu_a, 
\quad a \in \{1,2,3\}.
\end{equation}
Indeed, setting $u=-\mu_3$ we
find from~(\ref{eq-stat-exp})
\begin{equation}
\label{eq-solIIb}
\begin{split}
(\eta_{00}+ \mu_3)  K_0 &= -\frac{1}{2} \xi_0,\\
(\mu_1 -\mu_3)  K_1 &= 0,\\
(\mu_2 -\mu_3)  K_2 &= 0,\\
0 \cdot  K_3 &= 0.
\end{split}
\end{equation}
A solution of~(\ref{eq-solIIb}) which
also satisfies~(\ref{eq-ap-stat2})
and (\ref{eq-ap-stat3}) is
\begin{equation}
\label{eq-solIIbK}
K_0 =K_3 = \frac{-\frac{1}{2} \xi_0}{\eta_{00}+ \mu_3},
\quad
K_1 =K_2=0.
\end{equation}
In fact, any solution of~(\ref{eq-solIIb})
which respects~(\ref{eq-ap-stat2}) and (\ref{eq-ap-stat3})
can
be brought to the form~(\ref{eq-solIIbK}) by
a suitable basis change. This
holds, in particular, if there are
degeneracies of the eigenvalues $\mu_1, \mu_2$ with $\mu_3$.

Of course, we can have solutions 
of~\mbox{(\ref{eq-ap-stat1})-(\ref{eq-stat-exp})} analogous to~(\ref{eq-solIIbK})
for $u=-\mu_1$ and $u=-\mu_2$.
For values $u \notin \{-\mu_1, -\mu_2, -\mu_3 \}$
there are, clearly, no solutions 
of~(\ref{eq-ap-stat1})-(\ref{eq-stat-exp}).
Now we remember the ordering of the 
eigenvalues chosen in~(\ref{eq-sponEdiagord}).
The solution of~(\ref{eq-ap-stat1})-(\ref{eq-stat-exp})
with the largest Lagrange multiplier $u_0$ is, therefore,
given in~(\ref{eq-solIIbK}), corresponding to 
\begin{equation}
u=u_0=-\mu_3.
\end{equation}
According to theorem~3 of~\cite{Maniatis:2006fs}
the theory has the correct EWSB
and no zero mass charged Higgses if
and only if
\begin{equation}
\label{eq-ap-cond1}
u_0=-\mu_3>0.
\end{equation}
The vacuum solution is then given 
by~(\ref{eq-solIIbK}).
We know from the results
of~\cite{Maniatis:2006fs} that this gives
indeed the lowest potential value.
Here it is also straightforward to check
directly that for instance
the stationary point~(\ref{eq-regsol})
in the interior of the light cone
gives a higher potential value. 

Finally, it is clear that the
solution~(\ref{eq-solIIbK}) violates
the \CPg\ symmetry of type~$(i)$
spontaneously, since
\begin{equation}
\label{eq-ap-cond2}
- \tvec{K} \neq \tvec{K}.
\end{equation}
With~(\ref{eq-ap-stabstr}), (\ref{eq-ap-EWSBi})
and~(\ref{eq-ap-cond1}), (\ref{eq-ap-cond2}) we
have completed the investigation
of the stability and EWSB behaviour of
THDMs with \CPg\ invariance of type~$(i)$
and proven theorem~\ref{theorem-sponCP2}.


\section{Basis independent conditions for the absence of
spontaneous~\CPg~type~$(ii)$~violation}
\label{sec-independent-spon}

In this appendix we complete the proof of theorem~\ref{theorem-spCPgii}
by showing that the conditions~(\ref{cptransin-a})-(\ref{cptransin-d})
for the potential parameters together with the conditions
(\ref{eq-spcpi1})-(\ref{eq-spcpi3}) for the
vacuum expectation values
are equivalent to the existence of a basis with
\begin{align}
\label{eq-fullreal1}
\xi'_2&=0,\\
\label{eq-fullreal2}
\eta'_2 &=\eta'_{12}=\eta'_{23}=0,\\
\label{eq-fullreal3}
K_2'&=0.
\end{align}
Conditions \eqref{eq-fullreal1}-\eqref{eq-fullreal3} guarantee the existence of a
\CPg\ type (ii) invariance of both the potential and the vacuum expectation
values; see section~\ref{sec-subspii}.

We note first that~\eqref{eq-fullreal1}-\eqref{eq-fullreal3}
imply~(\ref{cptransin-a})-(\ref{cptransin-d}), 
see appendix~\ref{sec-independent},
as well as~(\ref{eq-spcpi1})-(\ref{eq-spcpi3}),
as can be seen immediately by direct insertion.

Now we show that from~(\ref{cptransin-a})-(\ref{cptransin-d}) 
and (\ref{eq-spcpi1})-(\ref{eq-spcpi3})
the existence of a basis satisfying~\eqref{eq-fullreal1}-\eqref{eq-fullreal3} follows.
We show this in two alternative ways. The first proof
reveals the number of geometric reflection symmetries for the different
cases. The second proof is more formal but also much shorter.

We discuss first the trivial case that the potential parameters
satisfy~(\ref{cptransin-a})-(\ref{cptransin-d}) and
the vacuum expectation value is the zero four-vector~$\fvec{K}=0$.
Then~(\ref{eq-spcpi1})-(\ref{eq-spcpi3}) are also
trivially satisfied. From theorem~\ref{theorem-CPgii} we
see that we can go to a basis where~\eqref{eq-fullreal1}
and \eqref{eq-fullreal2} hold. Since~$\tvec{K}=0$ in our case
we have also~$K_2'=0$, \mbox{q.e.d}.

Thus we can turn to the case that~\mbox{$\fvec{K} \neq 0$} 
which implies \mbox{$K_0 \neq 0$}; see~(\ref{eq-kinq}).
Then $\fvec{K}$ fulfills the stationarity condition
(see (91) and (96) of~\cite{Maniatis:2006fs})
\begin{equation}
\label{eq-statzeroapp}
\fvec{\xi}= -2 \left( \fmat{E} - u \fmat{g} \right) \fvec{K}
\end{equation}
where $u$ may be zero. For a theory
with the correct EWSB we have
$u=u_0= m_{H^\pm}^2/(2v_0^2)$,
see~(\ref{eq-stationarityzero4}),
but here we keep the discussion general and
do not assume this. In components 
we get from~(\ref{eq-statzeroapp})
\begin{equation}
\begin{split}
\label{eq-statzerocompapp}
\xi_0 &= -2 \big(\eta_{00} K_0 - u K_0 +\tvec{\eta}^\trans\tvec{K}\big),\\
\tvec{\xi}&= -2 \big(E \tvec{K}+ u \tvec{K}+K_0\tvec{\eta}\big).
\end{split}
\end{equation}

Consider now a potential with parameters
satisfying~(\ref{cptransin-a})-(\ref{cptransin-d}).
We may then choose a basis 
with $\tvec{\xi}'$, $\tvec{\eta}'$ and $E'$ 
of the form~\eqref{cptrans} by theorem \ref{theorem-CPgii}.
With a suitable rotation in the $1$--$3$ subspace we can 
diagonalise~$E'$. Then we have
\begin{equation}
\label{C6}
\tvec{\xi}'=
\begin{pmatrix}
\xi_1'\\ 0 \\ \xi_3'
\end{pmatrix},
\quad
\tvec{\eta}'=
\begin{pmatrix}
\eta_1'\\ 0 \\ \eta_3'
\end{pmatrix},
\quad
E'= \diag (\mu_1, \mu_2, \mu_3),
\end{equation}
\begin{align}
\label{eq-crossapp1}
\tvec{\xi}' \times \tvec{\eta}' &=
\begin{pmatrix}
0\\ \xi_3' \eta_1' - \xi_1' \eta_3'\\ 0
\end{pmatrix},\\
\label{eq-crossapp2}
\tvec{\xi}' \times E' \tvec{\xi}' &=
\begin{pmatrix}
0\\ (\mu_1 - \mu_3) \xi_1' \xi_3' \\ 0
\end{pmatrix},\\
\label{eq-crossapp3}
\tvec{\eta}' \times E' \tvec{\eta}' &=
\begin{pmatrix}
0\\ (\mu_1 - \mu_3) \eta_1' \eta_3' \\ 0
\end{pmatrix}.
\end{align}
In the basis of~(\ref{C6}) we fulfill
already~(\ref{eq-fullreal1}) and~(\ref{eq-fullreal2}).
It remains to be seen that also~(\ref{eq-fullreal3}) holds 
in this basis.
Let us first consider the case
%
\\[4mm]
$(a)\quad \tvec{\xi} \times \tvec{\eta} \neq 0\,:$
\nopagebreak\\[4mm]
This implies, of course,
$\tvec{\xi}' \times \tvec{\eta}' \neq 0$,
that is,
\begin{equation}
\xi_3' \eta_1' - \xi_1' \eta_3' \neq 0.
\end{equation}
If now~(\ref{eq-spcpi1}) holds
we get immediately
\begin{equation}
\label{eq-spcpcasea}
\begin{split}
&\left(\tvec{\xi}' \times \tvec{\eta}'\right)^\trans \tvec{K}'=0\\
&\Longrightarrow (\xi_3' \eta_1' - \xi_1' \eta_3') K_2' = 0\\
&\Longrightarrow K_2' = 0.
\end{split}
\end{equation}
Furthermore, we find from~(\ref{eq-crossapp2}),
(\ref{eq-crossapp3}) and~(\ref{eq-spcpcasea}) 
that~(\ref{eq-spcpi2}) and (\ref{eq-spcpi3}) are
automatically satisfied.
We summarise this case.
If $\tvec{\xi} \times \tvec{\eta} \neq 0$ the
only possible \CPg\ type~$(ii)$ symmetry
is the reflection on the plane spanned
by~$\tvec{\xi}$ and $\tvec{\eta}$ (see section~\ref{sec-explCP})
and this symmetry is respected by the vacuum
if and only if~(\ref{eq-spcpi1}) holds.
In this case~(\ref{eq-spcpi1}) implies
also~(\ref{eq-spcpi2}) and~(\ref{eq-spcpi3}).
This proves the case~$(a)$ of theorem~\ref{theorem-spCPgii}.

Next we consider the case
\\[4mm]
$(b) \quad \tvec{\xi} \times \tvec{\eta} = 0\,:$
\nopagebreak\\[4mm]
Then~(\ref{eq-spcpi1}) is trivially fulfilled.
Suppose first that~\mbox{$\tvec{\xi} \neq 0$}.
Then~$\tvec{\eta}$ is proportional to~$\tvec{\xi}$,
\begin{equation}
\tvec{\eta}= \lambda \tvec{\xi}.
\end{equation}
For the case of linearly dependent vectors~$\tvec{K}$  and $\tvec{\xi}$ 
we have
in particular in the basis defined by~(\ref{C6})
\mbox{$K'_2=0$} and (\ref{eq-fullreal3}) is proven. So
we may assume in the following that
$\tvec{K}$  and
$\tvec{\xi}$ are linearly independent.
Now we distinguish various subcases.
\\[4mm]
$(b.1)\quad \tvec{\xi}\times E \tvec{\xi} \neq 0\,:$
\nopagebreak\\[4mm]
The only reflection plane for a symmetry of the potential
is spanned by $\tvec{\xi}$ and $E \tvec{\xi}$ in this case.
We get from~(\ref{eq-crossapp2})
\begin{equation}
\left( \mu_1-\mu_3 \right) \xi_1' \xi_3' \neq 0
\end{equation}
and from~(\ref{eq-spcpi2})
\begin{equation}
\left( \mu_1-\mu_3 \right) \xi_1' \xi_3' K_2' = 0.
\end{equation}
This leads to~$K_2'$=0, q.e.d.
%
%
\noindent
\\[4mm]
$(b.2)\quad \tvec{\xi} \times E \tvec{\xi} = 0\,:$
\nopagebreak\\[4mm]
In this case we have
\begin{equation}
\label{eq-crossapp4}
\left( \mu_1-\mu_3 \right) \xi_1' \xi_3' = 0.
\end{equation}
Now we distinguish the different cases
for the eigenvalues of~$E$.
\\[4mm]
$(b.2.1)\quad \mu_1, \mu_2, \mu_3 \text{ all different:}$
\nopagebreak\\[4mm]
We get $\xi_1' \xi_3'=0$. If, for instance,
$\xi_1'=0$ the theory has two reflection symmetries
namely in this basis $R_1$ and $R_2$ (see (\ref{eq-genCPreflect})).
From~(\ref{eq-statzerocompapp})
we have
\begin{equation}
\begin{split}
0 &= -2 (\mu_1+u) K_1',\\
0 &= -2 (\mu_2+u) K_2'.
\end{split}
\end{equation}
Since we consider here \mbox{$\mu_1 \neq \mu_2$}
we must have either $K_1'=0$ or $K_2'=0$.
That is, at least one of the reflection symmetries
$R_1$ or $R_2$
is conserved by the vacuum.
In case~$K_1'=0$ we can by a change of basis
interchange the $1'$- and $2'$-components and
in this way achieve~$K_2'=0$, q.e.d. 
For $\xi_3'=0$ the argumentation is analogous, involving
$R_1$ and $R_3$.
%
\\[4mm]
$(b.2.2)\quad  \mu_1 = \mu_2 \neq \mu_3\,:$
\nopagebreak\\[4mm]
We get again from~(\ref{eq-crossapp4})
$\xi_1' \xi_3'=0$. For $\xi_3'=0$ the
argumentation is as in $(b.2.1)$. 
For~ \mbox{$\xi_3' \neq 0$} and
$\xi_1'=0$ 
we may perform a rotation 
around the $3'$-axis 
such that~$K'_2=0$ q.e.d.
Note, that~$E'$ is not
affected by this rotation since 
\mbox{$\mu_1=\mu_2$}.
In this case we
have reflection symmetry
on every plane containing the $3'$-axis, in particular
on the plane spanned by $\tvec{\xi}'$ and $\tvec{K}'$.
The reflection symmetry on this plane clearly is conserved
by the vacuum.
\\[4mm]
$(b.2.3)\quad \mu_2 = \mu_3 \neq \mu_1\,:$
\nopagebreak\\[4mm]
The argumentation is analogous to the case~$(b.2.2)$.
\\[4mm]
$(b.2.4)\quad \mu_1 = \mu_3 \neq \mu_2\,:$
\nopagebreak\\[4mm]
We can, by a rotation around the $2'$-axis, leaving
$E'$ diagonal, achieve
$\xi_1'=\xi_2'=0$, $\xi_3' \neq 0$.
Here~$R_1$ and $R_2$ are reflection symmetries.
Then~(\ref{eq-statzerocompapp}) gives
\begin{equation}
\begin{split}
0 &= (\mu_1+u) K_1',\\
0 &= (\mu_2+u) K_2'.
\end{split}
\end{equation}
Thus, either~$K_1'$ or $K_2'$ must be zero.
In case~$K_1'=0$ we can by a change of basis
interchange the $1'$- and $2'$-components and
in this way achieve~$K_2'=0$, q.e.d. 
%
\\[4mm]
$(b.2.5)\quad \mu_1 = \mu_2 = \mu_3\,:$
\nopagebreak\\[4mm]
There is reflection symmetry on all planes containing~$\tvec{\xi}'$,
in particular on
the plane spanned by~$\tvec{\xi}'$ and $\tvec{K}'$.
This reflection symmetry is obviously unbroken by the vacuum.
This proves theorem~\ref{theorem-spCPgii} for the case~$(b)$
if~\mbox{$\tvec{\xi} \neq 0$}.
For~\mbox{$\tvec{\eta} \neq 0$} everything runs analogously 
using~(\ref{eq-spcpi3}) instead of~(\ref{eq-spcpi2}).

\noindent
\\[4mm]
$(b.3)\quad \tvec{\xi} =\tvec{\eta} = 0\,:$
\nopagebreak\\[4mm]
In this case we have \CPg\ invariance of type~$(i)$.
There are then at least three \CPg\ type~$(ii)$ invariances.
We have here from~(\ref{eq-statzerocompapp})
\begin{equation}
\begin{split}
0 &= (\mu_1+u) K_1',\\
0 &= (\mu_2+u) K_2',\\
0 &= (\mu_3+u) K_3'.
\end{split}
\end{equation}
If not all~$\mu_a$ are equal this implies that at least 
one \mbox{$K_a'=0$}
\mbox{($a\in\{1,2,3\}$)}. 
By a change of basis we can always achieve that~\mbox{$K_2'=0$}, q.e.d.
If~\mbox{$\mu_1=\mu_2=\mu_3$} we have
reflection symmetry of the potential on any plane.
The reflection symmetries on all planes containing~$\tvec{K}'$ are
respected by the vacuum.
This completes the first proof of theorem~\ref{theorem-spCPgii}.

%
%
\begin{table}[t]
\begin{center}
\begin{tabular}{ll|c}
& parameter conditions\mbox{\;\;\;} & 
	 \; number of \CPg\ type~$(ii)$\\
& &
	\; reflection symmetries\\
\hline
\hline
$(a)$ & $\tvec{\xi} \times \tvec{\eta} \neq 0$ & $1$\\
\hline
$(b)$ & $\tvec{\xi} \times \tvec{\eta}=0$ & \\
& & \\
$(b.1)$ & $\tvec{\xi} \neq 0$, 
$\tvec{\xi} \times E \tvec{\xi} \neq 0$
& $1$\\
& $\tvec{\eta} \neq 0$, 
$\tvec{\eta} \times E \tvec{\eta} \neq 0$ & $1$\\
& & \\
$(b.2)$ & $\tvec{\xi} \neq 0$, 
$\tvec{\xi} \times E \tvec{\xi} = 0$ or & \\
&
$\tvec{\eta} \neq 0$, 
$\tvec{\eta} \times E \tvec{\eta} = 0$, & \\
&
eigenvalues of E: & \\
&
$\mu_1$, $\mu_2$, $\mu_3$ & \\
$(b.2.1)\,$ & $\mu_1$, $\mu_2$, $\mu_3$ all different
& $2$\\
$(b.2.2)$ &  $\mu_1=\mu_2 \neq \mu_3$
& $2$ or $\infty$\\
$(b.2.3)$ &  $\mu_2=\mu_3 \neq \mu_1$
& $2$ or $\infty$\\
$(b.2.4)$ &  $\mu_1=\mu_3 \neq \mu_2$
& $2$\\
$(b.2.5)$ &  $\mu_1=\mu_2=\mu_3$
& $\infty$\\
& & \\
$(b.3)$ &
$\tvec{\xi} = 0$, 
$\tvec{\eta} = 0$, & \\
&
$\mu_1$, $\mu_2$, $\mu_3$: &\\
&
all different
& 3\\
&
 at least 2 equal
& $\infty$\\
\end{tabular}
\end{center}
\caption{\label{t-typeiisym}
The \CPg\ type~$(ii)$ transformations are described by reflections
on planes.
The table lists the number of these
symmetries for a potential satisfying (\ref{cptransin-a})-(\ref{cptransin-d})
depending on the different cases for the parameters.
The vacuum is invariant under at least one of the symmetries
if and only if (\ref{eq-spcpi1})-(\ref{eq-spcpi3}) hold.
The numbering of the eigenvalues $\mu_1$, $\mu_2$, $\mu_3$ of $E$
is chosen such that $\mu_2=\eta'_{22}$ in a basis where
$\tvec{\xi}'$, $\tvec{\eta}'$ and $E'$ have the form~(\ref{cptrans}).
}
\end{table}

From the detailed discussion above we also found
the number of independent reflection symmetries, that is,
type~$(ii)$ \CPg\ transformations, which occur for 
the various cases. This is summarised in 
table~\ref{t-typeiisym} where it is always
supposed that the potential parameters
satisfy~(\ref{cptransin-a})-(\ref{cptransin-d}).


Now we present an alternative and more formal proof that
from~(\ref{cptransin-a})-(\ref{cptransin-d}) 
and (\ref{eq-spcpi1})-(\ref{eq-spcpi3})
the existence of a basis satisfying~\eqref{eq-fullreal1}-\eqref{eq-fullreal3} follows.
For the stationary point $\fvec{K}=0$, which leaves the electroweak
symmetry unbroken, the proof is trivial.
We shall now prove the statement for all other stationary points,
in particular for solutions with the required
EWSB.
We will use the fact that any stationary point $\fvec{K}\neq 0$
fulfills a stationarity condition of the form \eqref{eq-statzerocompapp}
with a specific value of $u$.
As a preparation we first show that certain additional invariants vanish.
Replacing $\tvec{\xi}$ in \eqref{eq-spcpi1} via the stationarity condition
\eqref{eq-statzerocompapp} we find
\begin{equation}
\label{eq-etaEKK}
\left( \tvec{\eta} \times (E \tvec{K}) \right)^\trans\, \tvec{K} = 0.
\end{equation}
This implies
\begin{equation}
\label{eq-xiEKK}
\left( \tvec{\xi} \times (E \tvec{K}) \right)^\trans\, \tvec{K} = 0,
\end{equation}
which can be seen by replacing $\tvec{\xi}$ via \eqref{eq-statzerocompapp}.
Next we show that
\begin{equation}
\label{eq-etaExiK}
\left( \tvec{\eta} \times (E \tvec{\xi}) \right)^\trans\, \tvec{K} = 0.
\end{equation}
If $\tvec{\eta}$ and $\tvec{K}$ are linearly dependent, \eqref{eq-etaExiK} follows
immediately. In the other case we replace $\tvec{\xi}$ in \eqref{eq-etaExiK}
by a linear combination of $\tvec{\eta}$ and $\tvec{K}$, which is possible by
\eqref{eq-spcpi1}. Using \eqref{eq-spcpi3} and \eqref{eq-etaEKK}, \eqref{eq-etaExiK}
follows.
Similarly we find
\begin{equation}
\label{eq-xiEetaK}
\left( \tvec{\xi} \times (E \tvec{\eta}) \right)^\trans\, \tvec{K} = 0,
\end{equation}
using \eqref{eq-spcpi1}, \eqref{eq-spcpi2} and \eqref{eq-xiEKK}.
The relation
\begin{equation}
\label{eq-EKExiK}
\left(E\tvec{K} \times (E\tvec{\xi})\right)^\trans\, \tvec{K} = 0
\end{equation}
follows after substitution of $E\tvec{K}$ via \eqref{eq-statzerocompapp}
from \eqref{eq-spcpi2} and \eqref{eq-etaExiK}.
Similarly we find
\begin{equation}
\label{eq-EKEetaK}
\left(E\tvec{K} \times (E\tvec{\eta})\right)^\trans\, \tvec{K} = 0
\end{equation}
using \eqref{eq-statzerocompapp}, \eqref{eq-spcpi3} and \eqref{eq-xiEetaK}.
We find
\begin{equation}
\label{eq-EKEEKK}
\left(\tvec{K} \times (E\tvec{K}) \right)^\trans\, E^2 \tvec{K} = 0
\end{equation}
by replacing $E\tvec{K}$ in the term $E^2 \tvec{K}$ via \eqref{eq-statzerocompapp}
since \eqref{eq-EKExiK} and \eqref{eq-EKEetaK} hold.

In the case that $\tvec{\xi}$ and $\tvec{\eta}$ are linearly independent,
we may choose a basis of the form~\eqref{cptrans} by theorem \ref{theorem-CPgii}.
From \eqref{eq-spcpi1} follows immediately that we have $K_2=0$ in this basis.

In the case that $\tvec{\xi}$ is a multiple of $\tvec{\eta}$ we
note that \eqref{eq-spcpi3}, \eqref{eq-etaEKK}, \eqref{eq-EKEEKK} and
\eqref{cptransin-d},
\begin{equation}
\begin{alignedat}{2}
\left(\tvec{K} \times \tvec{\eta}\right)^\trans\, E\tvec{K} &=0,&\quad
\left(\tvec{K} \times (E\tvec{K})\right)^\trans\, E^2\tvec{K} &=0,\\
\left(\tvec{K} \times \tvec{\eta}\right)^\trans\, E\tvec{\eta} &=0,&
\left(\tvec{\eta} \times (E\tvec{\eta})\right)^\trans\, E^2\tvec{\eta} &=0,
\end{alignedat}
\end{equation}
are equal to the explicit \CP\ conservation conditions
\eqref{cptransin-a}-\eqref{cptransin-d} if we replace
$\tvec{\xi}$ by $\tvec{K}$ in the latter.
Using the proof of theorem \ref{theorem-CPgii} we find that there is
a basis with $\eta'_2 = K'_2 = \eta'_{12} = \eta'_{23} = 0$ and thus also $\xi'_2 = 0$.

In the case that $\tvec{\eta}$ is a multiple of $\tvec{\xi}$ we use
\eqref{eq-spcpi2}, \eqref{eq-xiEKK}, \eqref{eq-EKEEKK} and
\eqref{cptransin-c},
\begin{equation}
\begin{alignedat}{2}
\left(\tvec{K} \times \tvec{\xi}\right)^\trans\, E\tvec{K} &=0,&\quad
\left(\tvec{K} \times (E\tvec{K})\right)^\trans\, E^2\tvec{K} &=0,\\
\left(\tvec{K} \times \tvec{\xi}\right)^\trans\, E\tvec{\xi} &=0,&
\left(\tvec{\xi} \times (E\tvec{\xi})\right)^\trans\, E^2\tvec{\xi} &=0.
\end{alignedat}
\end{equation}
Replacing $\tvec{\eta}$ by $\tvec{K}$ everywhere in the proof of
theorem~\ref{theorem-CPgii} we see that we can find a basis with
 $\xi'_2 = K'_2 = \eta'_{12} = \eta'_{23} = 0$ and thus also $\eta'_2 = 0$.
This completes the second proof of theorem~\ref{theorem-spCPgii}.

We compared our conditions~(\ref{eq-spcpi1})-(\ref{eq-spcpi3})
for absence of spontaneous \CP\ violation with those of theorem~4
in~\cite{Gunion:2005ja}.
The triple product in (\ref{eq-spcpi1}) equals $-(v/2)^4 \mIm J_1$ in
their notation, the other invariants in~\cite{Gunion:2005ja}
and our conditions have no one-to-one
correspondence.
However, we find complete agreement between
our conditions for absence of spontaneous \CP\ violation and those of~\cite{Gunion:2005ja}
taking into account the respective full set of equations, that is, including
the explicit \CP -conservation conditions and the stationarity equations.
This equivalence may be obtained via \emph{Groebner basis} computations.
Note however the comment in section~\ref{sec-subspii}
after theorem~\ref{theorem-spCPgii}
that ``absence of spontaneous CP violation'' is not quite
an appropriate formulation.
From the discussion of the case $(b)$ above and from
table~\ref{t-typeiisym} we see that, indeed,
a theory can have more than one \CPg\ type~$(ii)$ invariance.
One of these symmetries is always respected by the vacuum
if (\ref{eq-spcpi1})-(\ref{eq-spcpi3}) hold, but at the same time
others may be broken spontaneously.
We also compared our conditions~(\ref{eq-spcpi1})-(\ref{eq-spcpi3}) to
the corresponding conditions a)-c) in \cite{Nishi:2006tg}
and find agreement up to (\ref{eq-spcpi3}), which is not contained
in the latter set of criteria.
The condition~c) of~\cite{Nishi:2006tg} is no further
restriction since it is automatically fulfilled by 
the stationarity condition; see~\eqref{eq-xiEKK}.
Further, we do find examples
where omitting (\ref{eq-spcpi3}) matters,
that is examples satisfying the conditions of \cite{Nishi:2006tg} but
having spontaneous breaking of all \CP\ symmetries of the potential.

Let us now come back to Tab.~\ref{t-typeiisym} and
the cases of multiple \CPg\ symmetries of type~$(ii)$. 
Suppose we have in a theory two invariances of this type
denoted by \CPii\ and \CPiio.
Then the product $S \equiv \CPiio \circ \CPii$
is a conventional Higgs flavour symmetry. Indeed, from the
field transformation~(\ref{eq-genCPphi2}) we get
\begin{equation}
\begin{split}
\phi(x) &\xrightarrow{\CPii} \; U_\varphi \phi^*(x')\\
\phi(x) &\xrightarrow{\CPiio} \; U_\varphi' \phi^*(x')
\end{split}
\end{equation}
and
\begin{equation}
\phi(x) \xrightarrow{S} \; U'' \phi(x)\quad \text{with }
U'' =  U_\varphi' U_\varphi^* .
\end{equation}
Here $U_\varphi$, $U_\varphi'$ and $U''$ are all elements of $U(2)$.
Thus we see that in the cases of 2,3 or an infinite number
of \CPg\ transformations of type $(ii)$ as listed in
Tab.~\ref{t-typeiisym} there is a corresponding
number of Higgs flavour symmetries.
The possibility of a discrete ambiguity in the definition
of a generalised \CP\
transformation as a symmetry of the theory was also
noted in~\cite{Gunion:2005ja}.

Finally we discuss further the relation of 
the \CPg\ symmetries of type $(i)$ and $(ii)$.
Let us consider the generic case of a theory
where the Lagrangian is invariant under 
the type $(i)$ transformation,
that is the case (b.3) from 
Tab.~\ref{t-typeiisym}
with $\mu_1$, $\mu_2$, $\mu_3$
all different.
As we have shown in appendix~\ref{sec-ap-stability},
the \CPg\ transformation of type $(i)$ of the
fields is given in any basis by~(\ref{eq-genCPphieps}).
Clearly, we can consider this as product of
the standard \CPs\ transformation~(\ref{eq-simCP}) and
the Higgs flavour transformation induced by~$\epsilon$
\begin{equation}
\label{C30}
\begin{split}
\varphi_1(x) &\xrightarrow{} \phantom{+}\varphi_2(x),\\
\varphi_2(x) &\xrightarrow{} -\varphi_1(x).
\end{split}
\end{equation}
But note that in a given basis {\em neither}
this \CPs\ {\em nor} the transformation~(\ref{C30})
will in general be symmetries of the theory.
On the other hand we see from 
Tab.~\ref{t-typeiisym}  that a theory with \CPg\ invariance
of type $(i)$ automatically has
three \CPg\ invariances of type $(ii)$.
The latter are the reflections on the coordinate planes
in $\tvec{K}$ space only in the special basis where $E$ is diagonal.
There are also three
corresponding discrete Higgs flavour symmetries of
the type of a product of the \CPg\ symmetry
of type $(i)$ and one of type $(ii)$.
But this should be considered as a finding
a posteriori which is valid for the Higgs sector
of the theory taken in isolation. In the companion paper~\cite{Maniatis:2007de}
we find that for the complete theory, that is, the theory
including fermions, the \CPg\ symmetry of type $(i)$ does
in general not automatically imply invariance under the above
mentioned \CPg\ transformations of type $(ii)$. 
Thus, both from a conceptual point of view and from
exploring physical consequences, the \CPg\ transformations of
type $(i)$ and $(ii)$ should be considered independently
for their own sake.


\section{Theories with \CPg~type~$(ii)$~invariance}
\label{sec-ap-stability-ii}

Here we study the stability and EWSB
behaviour of models having a \CPg\ symmetry
of type~$(ii)$.
According to the discussion in section~\ref{sec-genCP}
we can then go to a basis~(\ref{cptrans}) where
$E'$ is already
partly diagonalised. By a change
of basis in the $1'$--$3'$ plane we can
diagonalise $E'$ completely without
changing the \CPg\ transformation
which is $R_2$~(\ref{eq-R2}) in this basis.
We then have 
\begin{equation}
E'= \diag (\mu_1, \mu_2, \mu_3 ),
\end{equation}
with $\eta'_{22}=\mu_2$ unchanged by the rotation
in the $1'$--$3'$ plane.
In the following all formulae
refer to this basis where we drop the prime
for ease of notation. Then we get
for the four-vector $\fvec{\xi}$ and
the \mbox{$4 \times 4$}~matrix $\fmat{E}$
defined in~(\ref{eq-fourpar})
\begin{align}
\label{eq-xicpgii}
\fvec{\xi}&=
\begin{pmatrix}
\xi_0\\
\xi_1\\
0\\
\xi_3
\end{pmatrix},\\
\label{eq-Ecpgii}
\fmat{E}&=
\begin{pmatrix}
\eta_{00} & \eta_1 & 0     & \eta_3\\
\eta_1    & \mu_1  & 0     & 0\\
0         & 0      & \mu_2 & 0\\
\eta_3    & 0      & 0     & \mu_3
\end{pmatrix}
\end{align}
with the potential
given by~(\ref{eq-Vfour}).
We must check the stability
of the potential.
Suppose this has been done, for instance by using theorem~1
of~\cite{Maniatis:2006fs}.

For a theory to have the correct
EWSB and no zero mass charged Higgs fields
the global minimum of $V$ must be
a solution of~(\ref{eq-ap-stat1})-(\ref{eq-ap-stat3}),
that is $\fvec{K}$ must be a
light-like four-vector.
The corresponding Lagrange multiplier $u_0$ must be positive
\begin{equation}
\label{eq-u0cond}
u_0>0\,,
\end{equation}
and it must be the largest Lagrange multiplier of all solutions
of~(\ref{eq-ap-stat1})-(\ref{eq-ap-stat3}).
According to theorem~3 of~\cite{Maniatis:2006fs}
these conditions are indeed not only necessary
but also sufficient for the determination of the global minimum
of an acceptable theory.

Now we can write out~(\ref{eq-ap-stat1})
in components. For~$K_2$ we find
\begin{equation}
\label{eq-sponCP-comp}
(\mu_2 + u) K_2 =0.
\end{equation}
Spontaneous violation of the \CPg\ type
$(ii)$ symmetry corresponding to $R_2$ in this basis means 
$K_2 \neq 0$.
Clearly, a solution 
of~(\ref{eq-sponCP-comp})
with $K_2 \neq 0$
requires
\begin{equation}
u=-\mu_2.
\end{equation}
This can correspond to the true vacuum solution only
if~$u=u_0>0$.
Thus, we find as necessary
condition for spontaneous
violation of this
\CPg\ type $(ii)$ symmetry from~(\ref{eq-u0cond})
and~(\ref{eq-sponCP-comp}) that
the eigenvalue~$\mu_2=\eta_{22}$ of
$E$ must be negative,
\begin{equation}
\mu_2=\eta_{22}<0.
\end{equation}
To prove that this \CPg\ symmetry is spontaneously
broken one still has to check if,
indeed,~(\ref{eq-ap-stat1})-(\ref{eq-ap-stat3}) have
a solution for $u=-\mu_2$ and
whether this is the solution with the
largest Lagrange multiplier $u=u_0$.
The above results are summarised in 
theorem~\ref{theorem-sponCPvac} 
in section~\ref{sec-subspii}.

Let us finally consider a potential with parameters
as in~(\ref{eq-xicpgii}), (\ref{eq-Ecpgii})
having (at least) two stationary
solutions on the light cone; see~(\ref{eq-ap-stat1})-(\ref{eq-ap-stat3}).
We suppose that the \CPg\ symmetry corresponding to the
reflection~$R_2$ in this basis
is respected by
one solution $\tilde{K}^\CPm$ with \mbox{$K^\CPm_2=0$} and violated by
the other solution $\tilde{K}^\CPviolm$ through \mbox{$K^\CPviolm_2\neq 0$}.
We denote the corresponding Lagrange multipliers by $u_\CPm$
and \mbox{$u_\CPviolm=-\mu_2$}.
Perturbing the \CPg\ conserving point by a small
amount \mbox{($0 < \varepsilon \ll 1$)} within the light cone
according to
\begin{equation}
\tilde{K}^\CPm \to \tilde{K}^\CPm
+K^\CPm_0 \begin{pmatrix}
  \sqrt{1 + \varepsilon^2} - 1\\
  0\\ \pm \varepsilon \\ 0 \end{pmatrix},
\end{equation}
we find for the potential value
\begin{equation}
V(\tilde{K}^\CPm) \to
V(\tilde{K}^\CPm) + (u_\CPm + \mu_2)\,\left(K^\CPm_0\right)^2\, \varepsilon^2
+ \mathcal{O}(\varepsilon^4)
\end{equation}
after employing the corresponding stationarity condition (\ref{eq-ap-stat1})
with $u=u_\CPm$.
Therefore,
the \CPg\ conserving point can only be a (local) minimum if \mbox{$u_\CPm + \mu_2 \geq 0$},
that is, if \mbox{$u_\CPm \geq u_\CPviolm$}.
From (123) in~\cite{Maniatis:2006fs} we know that a higher Lagrange
multiplier means a lower potential value.
To summarise,
if the potential has a \CPg\ conserving (local) minimum, there can
be no stationary points with lower values of the potential which
violate this symmetry.
This result was found before, see~\cite{Barroso:2005sm}
and references therein.
While the existence of a \CPg\ conserving light-like minimum implies
that the global minimum has these properties too,
there are cases with more than one \CPg\ conserving light-like minimum;
see Fig.~3 of~\cite{Maniatis:2006fs}.
Therefore, a determination of the
actual global minimum is still necessary in general.


\end{document}